%% file: eurosp-2020-template.tex
\documentclass[compsoc,conference,a4paper,10pt,times]{IEEEtran}
\IEEEoverridecommandlockouts
\usepackage{cite}
\usepackage{amsmath,amssymb,amsfonts}
\usepackage[ruled,vlined]{algorithm2e}
\usepackage{algpseudocode} 
\usepackage{subfig} 
\usepackage{graphicx}
\usepackage{textcomp}
\usepackage{bmpsize}
\usepackage{xcolor}
\usepackage{lipsum}
\usepackage{booktabs}
\usepackage{siunitx}
\usepackage{balance}

\usepackage[colorlinks=true,urlcolor=black]{hyperref}
\def\BibTeX{{\rm B\kern-.05em{\sc i\kern-.025em b}\kern-.08em
    T\kern-.1667em\lower.7ex\hbox{E}\kern-.125emX}}
\begin{document}

\newcommand\blfootnote[1]{%
  \begingroup
  \renewcommand\thefootnote{}\footnote{#1}%
  \addtocounter{footnote}{-1}%
  \endgroup
}

\title{Fall of Giants:\\How popular text-based MLaaS fall against a simple evasion attack}

\author{\IEEEauthorblockN{Luca Pajola}
\IEEEauthorblockA{\textit{Department of Mathematics} \\
\textit{University of Padua}\\
Padua, Italy\\
pajola@math.unipd.it}
\and
\IEEEauthorblockN{Mauro Conti}
\IEEEauthorblockA{\textit{Department of Mathematics} \\
\textit{University of Padua}\\
Padua, Italy\\
conti@math.unipd.it}
}

\maketitle

\begin{abstract}
The increased demand for machine learning applications made companies offer  Machine-Learning-as-a-Service (MLaaS). In MLaaS (a market estimated 8000M USD by 2025), users pay for well-performing ML models without dealing with the complicated training procedure. Among MLaaS,  text-based applications are the most popular ones (e.g., language translators).  
Given this popularity, MLaaS must provide resiliency to adversarial manipulations. 
For example, a wrong translation might lead to a misunderstanding between two parties.
In the text domain, state-of-the-art attacks mainly focus on strategies that leverage ML models' weaknesses.
Unfortunately, not much attention has been given to the other pipeline' stages, such as the indexing stage (i.e., when a sentence is converted from a textual to a numerical representation) that, if manipulated, can significantly affect the final performance of the application. 

In this paper, we propose a novel text evasion technique called ``\textit{Zero-Width} attack'' (ZeW) that leverages the injection of human non-readable characters, affecting indexing stage mechanisms.
We demonstrate that our simple yet effective attack deceives MLaaS of ``giants'' such as Amazon, Google, IBM, and Microsoft. Our case study, based on the manipulation of hateful tweets, shows that out of 12 analyzed services, only one is resistant to our injection strategy.
We finally introduce and test a simple \textit{input validation} defense that can prevent our proposed attack.
\end{abstract}


\input{Sections/Introduction}
\input{Sections/Background}

\input{Sections/Attack}
\input{Sections/CaseStudy}
\input{Sections/Results}

\input{Sections/Related_Works}
\input{Sections/Limitations}
\input{Sections/Conclusions}

\balance

\bibliographystyle{IEEEtran}
\bibliography{bibliography}

\end{document}

%% file: Sections/Introduction.tex
\section{Introduction}\label{sec.intro}

Without any doubt, machine learning applications had considerable success in the 2010s, finding space in different areas, from the automotive industry with autonomous vehicles~\cite{Sallab:2017:2470-1173:70} to the biomedical sector with brain tumor segmentation~\cite{Havaei_2017}. \blfootnote{This paper appears in the proceedings of the 6th IEEE European Symposium on Security and Privacy (EuroS\&P) 2021.}
The popularity of machine learning (ML) had a boost-up thanks to the increase of machines' computational power, making ML easily accessible to researchers and industrial developers, one of the significant obstacles of previous decades.
Even though ML nowadays is accessible to developers, we can find three main limitations in the deployment of ML solutions, due to the lack of: i) amount of data required to train a robust model, ii) amount of computational resources, and iii) machine-learning engineers with suitable expertise. For instance, we can consider the task of sentence language translation: in 2016, Google presented a translator based on Long Short-Term Memory (8 layers both encoder and decoder), trained over a parallel corpus\footnote{A parallel corpus is a dataset used for sequence-to-sequence tasks, where each sample has a source and a target. The goal of the model is to translate the source to the target.} of 26 million sentences (English-French)~\cite{wu2016googles}. Not only the difficulty of the model's architecture implementation (i.e., the choice of hyperparameters), but it requires an enormous amount of resources: the training procedure involved the use of 96 NVIDIA K80 GPU, with six days of computation.\par

The aftermath of such complexity is that real-world tasks are unlikely modeled with ML by companies or users without enough resources (i.e., computational power, data, ML engineers). 
To overcome this issue, the principal IT organizations (e.g., Amazon, IBM, Google, Microsoft) started developing solutions for common complex tasks (e.g., text analysis, optical character recognition) called Machine-Learning-as-a-Service (MLaaS), where users pay for a certain amount of queries. 
In this way, for example, companies that require to analyze documents can use advance and well-performing techniques at an affordable price without caring for the complex training process. 
MLaaS had a discrete success, and in 2019 this market was valued 1.0 billion USD, with an estimation of 8.48 billion USD by 2025~\cite{mordor}. \par

The rapid growth of ML in real case applications also attracted the security community. 
Researchers started asking whether users can maliciously affect ML-applications decisions: this area is called \textit{Adversarial Machine Learning}~\cite{10.1145/1014052.1014066}. In particular, several proposed attacks show the feasibility of affecting at test time ML algorithms by adding small and unnoticeable perturbation to the input data~\cite{Biggio_2013},~\cite{goodfellow2014explaining},~\cite{papernot2016transferability}. 

In this paper, we focus on the text-domain, where the addition of malicious perturbation is translated with the modification of text through various techniques (e.g., misspelling, typos, word addition). 
The primary constraint of attacks in the text-domain is the ``readability preservation’’, i.e., a human being can understand the meaning of modified sentences.
The reasons behind the limitation mentioned above are deducible: let us consider the sentiment classification of music reviews (i.e., positive/negative reviews), where the adversary goal is to have positive classifications for negative sentences. The original sentence could be ``I hate this album'', and its malicious counterpart ``I hA.XYztXaeX this album''. 
While the original sentence classification is expected to be negative, its counterpart might not; however, from a human point-of-view, the malicious sentence is not readable, and thus the adversarial sample loses its semantic meaning. 
So far, several works have proposed adversarial techniques in the text-domain, combining complex algorithms to find a trade-off between the effectiveness of the attack and the readability of produced malicious sentences. For instance, in~\cite{samanta2017crafting}, the authors identify the importance of the words of a given sentence for the target model, and replace them with sophisticated linguistic strategies. 
In~\cite{8424632}, the authors proposed \textit{DeepWordBug}, a process that first uses a scoring function to identify critical tokens for the target model and then applies character transformations to minimize the number of modifications. For an in-depth overview of state-of-the-art adversarial machine learning in text-domain, we suggest~\cite{10.1145/3374217}.

\begin{figure*}
    \centering
    \includegraphics[width=1.0\linewidth]{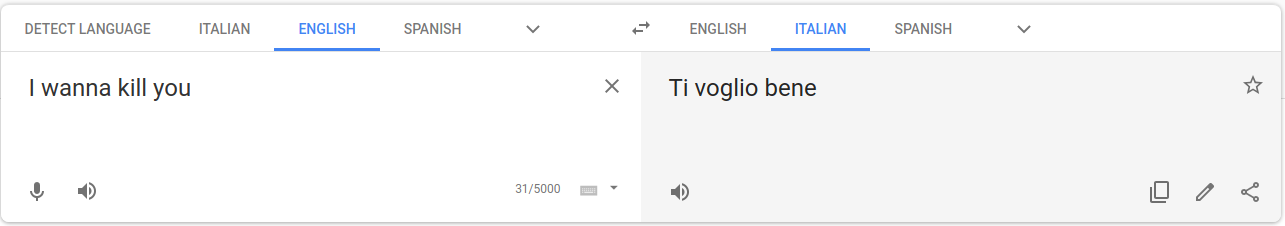}
    \caption{Zero-Width (ZeW) on a real-life scenario: Google Translate. The translated sentence means ``I love you''.}
    \label{fig:g_translate}
\end{figure*}

\paragraph{\textit{Contributions}}
Motivated by the common assumption of ``readability preservation’’, we investigate a novel evasion technique that guarantees full readability and attack effectiveness. 
Our technique, called ``\textit{Zero-Width}'' (ZeW), injects malicious UNICODE characters often used in text steganography strategies. 
These characters are called \textit{zero-width space}, and their effect is that, when printed, they have zero-width, resulting invisible from a human being's perspective. 
In one of our attack scenarios, we attack the popular web application Google Translate\footnote{\url{https://translate.google.com}.} on the English-Italian task. Figure~\ref{fig:g_translate} shows an example of wrong translation, where the original sentence ``I wanna kill you'' is translated as ``ti voglio bene'', which means ``I love you''. It is curious to notice that the input section has 31 characters (Figure~\ref{fig:g_translate}, left side), while the sentence should contain only 16. 
In contrast to the state-of-the-art, ZeW does not require any assumption of the target model, and the readability constraint is relaxed. 
Moreover, so far, most of the proposed attack strategies aim to leverage the learning strategies’ weaknesses (e.g., model architectures); however, a ML application is composed of several stages (pipeline), where the ML model is only one of them. In this work, we aim to attack and disrupt the ``indexing-stage'' (see Section~\ref{sub.textpipeline}), which is the step that converts a sentence from the textual representation to a numerical one. To the best of our knowledge, not much attention has been put to find possible weaknesses to the entire text pipeline.  

In this paper we aim to understand the following: i) the effect of ZeW attack on different types of indexing strategies, and ii) if commercial solutions are vulnerable to ZeW attacks. 
We conduct our experiments through a case study of a possible ZeW attack application: hate speech manipulation. We designed a simple injection strategy that, given a hateful sentence, identifies negative words and injects malicious characters in two possible fashions: i) \textit{Mask1}, where only one malicious character is inserted in the middle of the word and ii) \textit{Mask2}, where one malicious character is inserted between each character of the word. We tested this strategy over popular text MLaaS provided by Amazon, Google, IBM, and Microsoft, without having prior knowledge of the target models.
The analysis aims to understand which services can be affected by ZeW, and the magnitude of the vulnerability. 
Our experiment shows that 11 out of 12 MLaaS are vulnerable to the proposed attack. We further introduce a simple countermeasure approach that can prevent ZeW. The purpose of this work is to emphasize the importance of studying the security of machine learning pipelines in all of their stages. Due to the gravity of ZeW, at the time of submission, all of the companies (i.e., Amazon, Google, IBM, and Microsoft) are informed.\par
Our contributions can be summarized as follows.
\begin{itemize}
    \item We propose a novel text evasion strategy called \textit{Zero-Width} (ZeW) that affects the indexing stage of text pipelines.
    \item We show the effect of ZeW over Machine-Learning-as-a-Service developed by Amazon, Google, IBM, and Microsoft. Out of 12 tested services, 11 show vulnerabilities (8 strongly affected).  
    \item We propose a countermeasure to ZeW that can be easily integrated in every text ML-based pipeline.
\end{itemize}

\paragraph{\textit{Paper Organization}}
The manuscript is organized as follows. In Section~\ref{sec.prel} we first briefly introduce the basic concepts required to fully understand the rest of the paper. Motivations, theoretical perspective, and countermeasure of ZeW are described in Section~\ref{sec.zwspa}. We then move in Section~\ref{sec.cs} with the implementation of ZeW in a real case scenario, the hate speech manipulation, followed by a discussion of the attack results in a controlled environment first (Section~\ref{sec.res_indoor}), followed by MLaaS (Section~\ref{sec.results}).
In Section~\ref{sec.rel_work} we summarize state-of-the-art attacks targeting models of MLaaS. 
We conclude with the limitations of the proposed attack in Section~\ref{sec.limitations}, followed by considerations and discussions of the possible implications of our results (Section~\ref{sec.end}). 

%% file: Sections/Background.tex
\section{Background and Preliminaries}\label{sec.prel}
This section presents the preliminary concepts required for the rest of the paper. Section~\ref{sub.textpipeline} presents an overview of the standard pipeline in Natural Language Processing (NLP) applications, followed by an introduction to the adversarial machine learning theory in Section~\ref{sub.aml}, and, finally, Section~\ref{sub.text_aml} describes keys and challenges of adversarial machine learning in the text-domain.

\subsection{Text Pipeline}\label{sub.textpipeline}
\begin{figure}[!ht]
    \centering
    \includegraphics[width=1.0\linewidth]{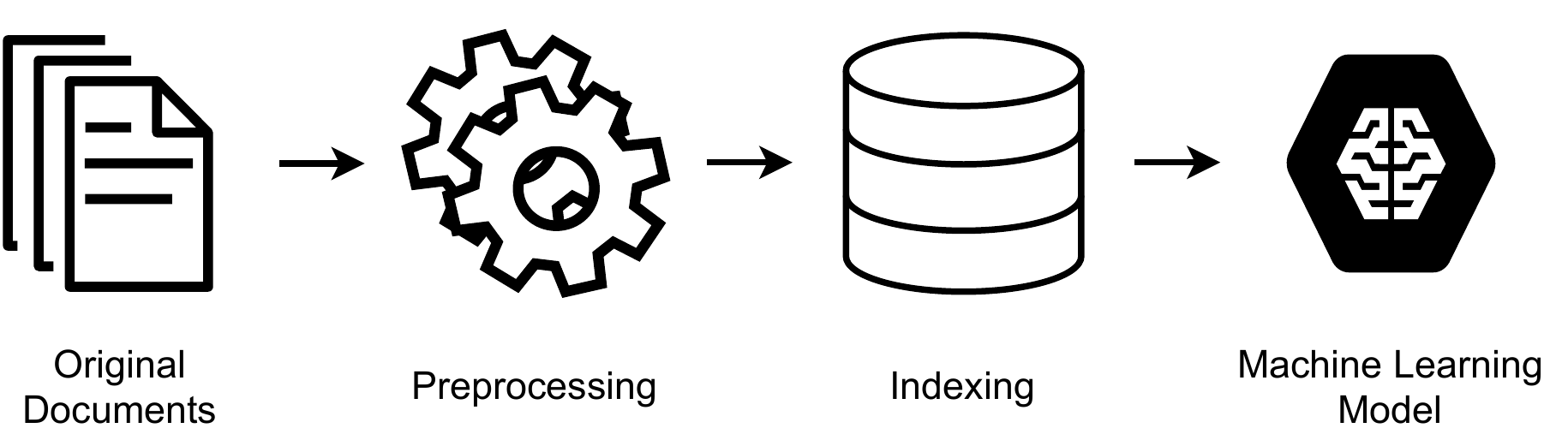}
    \caption{Machine Learning pipeline in Natural Language Processing.}
    \label{fig:text_pipeline}
\end{figure}

ML-based applications on text-domain follow a common pipeline, as described in~\cite{10.1145/3374217, Korde}, and shown in Figure \ref{fig:text_pipeline}. 
The pipeline consists of the following components: original documents, preprocessing, indexing, and machine learning model.
    \paragraph{\textit{Original Documents}}Collection of the corpus of documents to analyze. The origin of these documents can differ, such as text files, PDFs, or HTML web pages.
    \paragraph{\textit{Preprocessing}}Set of mechanisms that prettify documents, with the removal of useless information (e.g., TAG, format controls). This stage can involve different techniques, such as \textit{tokenization}, where sentences are decomposed in lists of words, \textit{stopword removal}, where common words (and meaningless) are removed (e.g., articles), and \textit{stemming}, where words are converted in their root form (e.g., books $\xrightarrow{}$ book). 
    \paragraph{\textit{Indexing}}The mechanism that converts the symbolic representation of a document/sentence into a numerical vector. At training time, a vocabulary $V$ of the possible representations (word/character level) is defined. The vectorial representation is usually handled in three possible ways:
    \begin{itemize}
        \item \textit{Word-count encoding}. Each document is represented as a vector of words occurrences. For example, given the sentences $s_1=$ ``hello there'' and $s_2=$ ``hello hello''', a vocabulary $V=[hello, there]$, the sentences are represented as
        \begin{align*}
            &s_1=[1, 1],\\&s_2=[2,0],
        \end{align*}
        where the numbers represents the number of occurrences of the correspondent index in the vocabulary (i.e., ``hello'' in position 0, ``there'' in position 1). A variant of the word count often use the Term Frequency-Inverse Document Frequency (TF-IDF); this encoding tries to capture the importance of a word in the document given a collection of documents. 
        \item \textit{One-hot encoding}. This encoding represents a document as a list of vectors (one per word/char in the document). Given the previous example, the sentences are represented as
        \begin{align*}
            &s_1=[[1, 0], [0,1]],\\&s_2=[[1, 0], [1,0]].
        \end{align*}
        \item \textit{Dense encoding}. In this category, we find word embeddings, powerful vectorial representations of words~\cite{mikolov2013efficient, 10.5555/2999792.2999959}. Here, each word is represented as a vector of real numbers (abstract representation). Dense representations can be pre-trained or trained end-to-end (e.g., using Language Models). 
        For example, given the words ``dog'', ``cat'' and ``hello'', the dense representation indicates that the word ``dog'' is spatially closer to ``cat'' rather than ``hello''.  
    \end{itemize}
    During the indexing phase, the pipeline needs to deal with unrecognized items (word/character level), i.e., items out-of-vocabulary (OOV). There are two possible ways to deal with it: i) discarding them from the analysis or ii) mapping them to a special token ``UNK''. In the latter case, the special token has a proper representation, based on the indexing strategy (i.e., word-counting encoding, one-hot encoding, dense encoding). 
    The problem of unrecognized items is well-established in NLP, where the frequency of items in a dataset usually follows long-tail distributions. To reduce the complexity of the problem, the standard approach is to maintain small vocabularies with the most frequent items~\cite{chen-etal-2019-large}. In Natural Machine Translation tasks this could be a problem, where all of the OOVs are mapped to a single token ``UNK''. For example, we can consider the translation task from English to a target language of the sentence ``Liam meets Noel''. Likely, both proper names are not present in the vocabulary, and thus they are mapped to the same token (i.e., ``UNK meets UNK''), losing the name information in the target language. A standard approach, proposed in~\cite{luong-etal-2015-addressing}, consists of using placeholders to map rare items with  unique pointers (e.g., ``UNK1 meets UNK2'', where UNK1 = Liam and UNK2 = Noel) with a final name replacement in post-processing.   
    \paragraph{\textit{Machine Learning Model}}The ML model used for the task. The set of models vary from simple architectures (e.g., Logistic Regression, Random Forest), to Neural Networks (NN) and Deep Neural Networks (DNN). In the latter case, we can find variants of Recurrent Neural Networks (RNN) such as Long Short-Term Memory (LSTM)~\cite{doi:10.1162/neco.1997.9.8.1735} or Gated Recurrent Units (GRU)~\cite{chung2014empirical}.

ZeW aims to attack and disrupt the pipeline by affecting the indexing stage, affecting the ML-model performance.

\subsection{Adversarial Machine Learning \& Application Security}\label{sub.aml}

Adversarial Machine Learning (AML) is the discipline that studies the security of ML algorithms~\cite{Huang:2011:AML:2046684.2046692, Biggio_2013}. 
In literature, we can find several classes of attacks. 
For example, the model \textit{evasion attack}, where the adversary defines an \textit{adversarial example} with the aim to affect target ML predictions~\cite{Biggio_2013, goodfellow2014explaining}. Formally, given a target model $\mathcal{F}$, an input sample $x$, the adversary aims to find a small  \textit{perturbation} $\epsilon$ in such a way that $\mathcal{F}(x) = y_i$ and $\mathcal{F}(x + \epsilon) = y_j$, where $y_i \neq y_j$. 
Another popular attack is the \textit{poisoning attack}: if the attacker has access to the training data, he/she can inject malicious samples that affect the model performance \cite{10.5555/3042573.3042761}. 
A variant of this attack is called \textit{trojan/backdoor attack}, where the attacker does not influence the model performance, but instead creates a backdoor in the model. At test time, the attacker triggers this backdoor, with results similar to the evasion \cite{gu2017badnets, 10.1145/3394486.3403064}.

Nevertheless, ML applications contain not only ML algorithms but also additional steps such as preprocessing and transformations. For example, in Section~\ref{sub.textpipeline} we described an overview of standard text pipelines. 
Therefore, the concept of \textit{applications security} must assess each component that defines such pipelines.

Despite its importance, not a lot of attention is given to this broader area of application security.  
Examples of software security-related vulnerabilities are presented in~\cite{8424643}, where the authors disclose a set of attacks (e.g., denial of service) related to popular ML frameworks as Caffe and Tensorflow. 
Similarly, in~\cite{236348} the authors introduce the \textit{Camouflage Attack}, where malicious images change their semantic meaning after scaling them. Such attack can be exploited in computer vision (CV) applications, where image scaler algorithms are usually employed upstream of CV pipelines.

We remark that adversarial machine learning techniques aim at machine learning models, while our proposed attack ZeW exploits vulnerabilities of \textit{preprocessing} and \textit{indexing engine} algorithms of text pipelines (see Section~\ref{sub.textpipeline}).

\subsection{Challenges of adversaries in Text-Domain}\label{sub.text_aml}

While AML gained popularity in Computer Vision (CV) from its early stages, only in recent years researchers moved onto the NLP domain. As identified by~\cite{10.1145/3374217}, three major aspects differentiate AML in NLP from CV.
\begin{itemize}
    \item \textit{Input Domain}. While images are defined in a continuous space (e.g., RGB matrix), sentences are discrete and represented as a list of symbols. It implies that the meaning of perturbation that we want to add changes its nature. For example, in CV the perturbation is defined as a matrix of values to sum up to the original image. This is not possible in NLP since there is no meaning in adding an integer to a word (e.g., ``dog’’ + 1).
    \item \textit{Human Perception}. From a human point of view, perturbations in CV are difficult to perceive, since the modifications are at a pixel level. Vice-versa, on text, small changes are easily detectable by both human beings and machines (e.g., spell checkers).
    \item \textit{Semantic}. From the semantic point-of-view, the addition of a perturbation into an image rarely changes its meaning. In NLP, the modification/addition/removal of a character/word may lead to a completely different meaning of the sentence (e.g., ``I hate you’’, ``I ate you’’). 
\end{itemize}

As a consequence, state-of-the-art attacks on NLP are either CV-algorithms adapted to face NLP challenges, or novel solutions designed from scratch. 


In this work, we are mainly interested in the \textit{evasion attack}. As previously introduced, our goal is to define a perturbation that influences the target model while preserving the semantic and readability of the sentence. A small amount of perturbation can guarantee a correct human perception; for example, in~\cite{leet}, authors show the human resistance to leet speech, e.g., ``R34D1NG W0RD5 W1TH NUMB3R5''. 
The choice of the measurement is not trivial as in CV, where spatial distance metrics between the original sample $x$ and the malicious $x'$ are used. As stated in~\cite{10.1145/3374217}, we can measure the perturbation in different ways, such as norm-based distances for dense representations, or edit-based measurement, which identify the number of changes required by making $x'$ equal to $x$.


%% file: Sections/Attack.tex
\section{Zero-Width Attack}\label{sec.zwspa}
In this section, we present the Zero-Width attack (ZeW).
We first introduce in Section~\ref{sub.motivations} the motivations and the intuition that drives our investigation. In Section~\ref{sub.tp} we describe how ZeW can affect different NLP pipelines. We conclude with Section~\ref{sub.counter} by describing a countermeasure to our proposed attack.

\subsection{Motivations}\label{sub.motivations}
Three main motivations guide our investigation.
\begin{enumerate}
    \item \textit{UNICODE representation}. Most NLP tools allow the use of UNICODE characters. This is essential, especially for the analysis of web text. For example, on Social Networks, non-ASCII characters are often used (e.g., emoji).
    \item \textit{Readability Preservation}. The attack strategy should apply fewer modifications as possible to maintain the sentence readability.
    \item \textit{Indexing stage vulnerabilities}. To the best of our knowledge, most of the attack strategies aim to leverage ML-models' weaknesses, while little attention has been put to the security of other stages of the text ML pipeline, such as the indexing stage (see Section~\ref{sub.textpipeline}).
\end{enumerate}

We asked ourselves if it exists a technique that allows us to relax the constraint of the number of modifications to malicious sentences, allowing us to focus only on the disruption of target models' performance. 
We found the answer in the \textit{steganography} discipline, which is the ``art of hiding secret messages into plain sources''~\cite{Bennett04linguisticsteganography}. In the UNICODE representation, there are characters whose width is zero, i.e., when printed, they are invisible, and human beings cannot perceive them. 
Some examples of these characters are \textit{zero-width space} (U+200B) and \textit{zero-width joiner} (U+200C). These allow us to insert an arbitrary number of ``invisible'' characters in a given sentence. 
Thanks to this particular property, we can forget to consider the problem of readability preservation since sentences semantic is intact. 
The presence of zero-width characters allows an attacker to affect the decision of the indexing stage (see Section~\ref{sub.tp}).
We identify in total 24 malicious characters\footnote{https://github.com/pajola/ZeW/blob/main/ZeW.py}. 

In cybersecurity, we can find the usage of zero-width characters in different ways. For example, in~\cite{7346813} the authors use zero-width characters in the communication protocol of a botnet, ELISA; here, the botmaster secretly communicates through public posts with the zombies over social networks such as Facebook. In late 2018, the security team AVANAN discovered a phishing method against Office 365, bypassing Microsoft's security mechanisms~\cite{avanan}; in this attack, hackers used zero-width characters in the middle of malicious URLs, evading Microsoft's detection mechanisms. While in security zero-width characters are a known threat, to the best of our knowledge, we are the first to explore their effect in the adversarial machine learning context.

\subsection{Theoretical Perspective}\label{sub.tp}
Zero-width characters give us the power to break the intra-relationship between the characters of a given sentence. Let's represent from now on zero-width characters with the symbol ``\$''. We can recall the example reported in Section~\ref{sec.intro} ``I hate this album''; the malicious version ``I h\$a\$t\$e this album'' appears identical to the original sentence from a human point of view, while different from a machine perspective. Figure~\ref{fig:g_translate} presents a real example of zero-width characters. We can notice that the malicious sentence appears legitimate.

In Section~\ref{sub.textpipeline} we described possible numerical representations of a given sentence (\textit{indexing} stage). We now explain how ZeW can affect those representations.
\begin{itemize}
    \item \textit{Word-based representations}. In word-based representations, a sentence can be seen as a temporal vector $s = t_0 + t_1 + ... + t_n$, where $t_i$ is the token (i.e., word, punctuation symbol) at the time $i$, and $n$ is the length of the tokenized sentence (see Section~\ref{fig:text_pipeline}, \textit{preprocessing} stage). Here, it is unlikely that words containing ``\$'' are present in the vocabulary $V$.
    Two possible scenarios can occur.
    \begin{itemize}
        \item Unrecognized words are mapped to special tokens (e.g., placeholders, ``UNK''). It is likely that unpoisoned words and ``UNK'' have different meanings and effects to target models, since they appear with a different representation. For example, the sentence ``I h\$a\$t\$e this album'' is represented as ``[I, UNK, this, album]''.
        \item Unrecognized words are discarded from the analysis, with a consequence of loose of expressiveness of the malicious sentence. For example, the sentence ``I h\$a\$t\$e this album'' is represented as ``[I, this, album]'' (the word ``h\$a\$t\$e'' is discarded). In this case, the target model analyzes only the remaining sentence. Potentially, by adding one zero-width character per token, the resulting sentence will be empty. 
    \end{itemize}
    
    \item \textit{Character-based representations}. In char-based representations, a sentence can be seen as a temporal vector $s = t_0 + t_1 + ... + t_n$, where $t_i$ is the character at position $i$, and $n$ is the total number of characters that compose the sentence. As in the previous case, two possible scenarios can occur.
    \begin{itemize}
        \item Unrecognized characters are mapped to the special tokens (e.g., placeholders, ``UNK''), resulting in an addition of noise in the vectorial representation. For example, the word ``h\$a\$t\$e'' is represented as ``[h, UNK, a, UNK, t, UNK, e]''.
        \item Unrecognized characters are discarded from the analysis. In this case, the poisoned sentence coincides with the original sentence. The attack has no effect in this scenario. For example, the word ``h\$a\$t\$e'' is correctly represented as ``[h, a, t, e]''.
    \end{itemize}
    
\end{itemize}

In general, ZeW leads to an increase in noise or reduction of information in the sentence representation. The attack can be seen as an \textit{injection} attack, where malicious characters are injected into target sentences. Potentially, an attacker can insert an arbitrary number of ``\$'' on malicious sentences, without any constraint. This gave us the capability of not considering the perturbation measurement described in Section~\ref{sub.text_aml}. 


To the best of our knowledge, injection strategies using ZeW characters can be further optimized with target ML-models, resulting in the following adversarial attacks:
\begin{itemize}
    \item \textit{Evasion}. ZeW characters can be optimally inserted in target sentences to affect ML models' decisions. 
    \item \textit{Poisoning}. If the adversary has access to the training data, the addition of malicious samples could lead to a noisy dataset, decreasing the overall performance. 
    \item \textit{Trojan}. If the adversary has access to the training data, he/she can inject a rare sequence of zero-width characters in a small portion of the dataset and let the model overfit over them. At test time, the trojan is triggered by samples containing that specific sequence.
\end{itemize}
The definition of such adversarial attacks is out of the scope of the paper. 

\subsection{Countermeasure}\label{sub.counter}
Overall, ZeW is an injection attack that influences the indexing stage (see Section~\ref{sub.textpipeline}), with consequences in the following steps (i.e., machine learning algorithms). 
ZeW leverages peculiar properties of UNICODE representation, which contains non-printable characters. In the security field, injection attacks are a standard and well-known problem~\cite{10.1145/1111037.1111070}. 
A typical example is the SQL injection, where the definition of malicious input can damage the target database structure and destroy its contents. Injections can be severe, especially when users are allowed to insert arbitrary input used for critical operations. 
Similarly, MLaaS offer users to interact with ML-models through APIs. It is thus essential to have mechanisms that control any input feeding the models, placed at the \textit{preprocessing stage}; these are also called sanitization or input validation mechanisms. Regarding ZeW, a simple filter that rejects malicious sentences containing non-printable characters is enough. Similarly, the sanitizer can just discard the malicious characters.





%% file: Sections/CaseStudy.tex
\section{Case Study: Hate Speech Manipulation}\label{sec.cs}
In this section we introduce the disignated case study (Section~\ref{sub.cs_over}), followed by the injector algorithm description (Section~\ref{sub.alg}).

\subsection{Overview}\label{sub.cs_over}
We test and evaluate ZeW on text Machine-Learning-as-a-Service provided by popular companies such as Amazon, Google, IBM, and Microsoft. These services vary from sentiment analyzer to language translators. Our idea is to test some of the most popular ML-based text services to understand how many applications can be affected by the attack. 

As a case study, we analyze the \textit{hate speech manipulation}, a topic that raised the interest of a broad area of researchers in the last years~\cite{waseem-hovy-2016-hateful, schmidt-wiegand-2017-survey}. Our goal is to understand how zero-width characters affect the outcomes of different MLaaS. We consider the attack successful if the injection of zero-width characters affects in some way the performance of a target model. We are also interested to understand the magnitude of the vulnerability.  
In our opinion, this is a likely scenario where a malicious user aims to offend a target victim without being detected, since it is known the problem of malicious interaction between users and Artificial Intelligence systems. A famous example is the Microsoft chatbot Tay, which becomes hateful after a poisoning attack of a group of users~\cite{WOLF20171}. 


\subsection{Manipulation algorithm}\label{sub.alg}
In this work, we aim to define a simple yet effective strategy using ZeW attack. Simple and non-optimal attacks have been shown to be effective in~\cite{8835391}, where cybercriminals evade sexually explicit content detection with simple image transformations (e.g., random noise addition).  

In our attack, we assume that hateful sentences contain a negative part-of-speech, as shown in Figure~\ref{fig:vader} on the \textit{Real} corpus. We thus want to understand how the performance of the tested MLaaS are affected when ``deleting'' such negative parts. To do so, we designed a simple injection strategy that, given a sentence, identifies negative words (i.e., words with negative polarity scores) and injects on them zero-width characters. In the experiment, we inject zero-width characters in two possible fashions.
\begin{enumerate}
    \item \textit{Mask1}. Only one random Zero-Width SPace character is injected in the middle of the target word (e.g., $hate \longrightarrow ha\$te$).  
    \item \textit{Mask2}. Multiple random Zero-Width SPace characters are injected, one between each character (e.g., $hate \longrightarrow \$h\$a\$t\$e\$$).
\end{enumerate}
The idea behind these two strategies is to measure the impact of ZeW with different levels of injection. Algorithm~\ref{Alg.1} shows the overall attack strategy.
To identify negative words we use \textit{VaderSentiment}, a free sentiment analyzer tool available for Python~\cite{vader}. The code of the injector is available on GitHub\footnote{https://github.com/pajola/ZeW}.

\begin{algorithm}
\SetAlgoLined
\SetKwInOut{Input}{input}
\SetKwInOut{Output}{output}
\Input{An original sentence $s$ and the type of injection mask $m$}
\Output{A poisoned sentence $s_{pois}$}
$tokens=Tokenizer(s)$\\
$N_{tok} = length(t)$\\
$i = 0$
$s_{pois} = []$
\While{$i < N_{tok}$}{
    $t = tokens[i]$\\
    $t_{stem} = Stem(t)$\\
    $t_{sent} = Sentiment(t_{stem})$\\
    \If{$t_{sent}\;is\;negative$}{
        $t_{pois} = Injector(t, m)$\\
    }
    $s_{pois}.add(t_{pois})$\\
    $i = i + 1$\\
}
$s_{pois} = Join(s_{pois})$
\caption{HS-Manipulation}
\label{Alg.1}
\end{algorithm}

%% file: Sections/Results.tex
\section{Results on Controlled Environments}\label{sec.res_indoor}
In this section, we evaluate the impact of the ZeW injection strategy presented in Section~\ref{sec.cs} over different machine learning models and indexing techniques. In Section~\ref{sub.res_indoor_sett} we first present the experimental settings, followed by result discussions in Section~\ref{sub.res_indoor_res}. 

\subsection{Experimental Settings}\label{sub.res_indoor_sett}
Algorithm~\ref{Alg.1} aims to reduce the negative part-of-speech of a given sentence. We thus decide to understand the impact of ZeW injection strategy over a binary classification task: the sentiment classification. The task consists of predicting whether a sentence is positive or negative. For the experiments we use the \textit{Sentiment140 dataset}~\cite{Sentiment140}. The dataset contains positive and negative tweets (800K per class), for a total of 1.6M of labeled tweets. We then randomly split the corpus into a training (70\%), validation (10\%), and testing partition (20\%). 
We evaluate two types of ML algorithms:
\begin{itemize}
    \item \textit{SGDClassifier}. This is a linear classifier. We use Scikit-Learn~\cite{scikit-learn} implementation. The model is built on top of a \textit{TfidfVectorizer}, i.e., an engine that converts raw documents into TF-IDF representations.  
    \item \textit{Recurrent Neural Network Classifier}. We deploy standard RNN-based classifiers using an embedder, followed by a two layers GRU and a final linear layer. The model is deployed using Pytorch~\cite{NEURIPS2019_9015}. 
\end{itemize}
Each model is trained over different variants of text representation i.e., character and word-based. The SGDClassifier implements two different combinations: character ngrams defined in the range $[1, 5]$, and word ngrams, defined in the range $[1, 3]$. For example, the range $[1, 2]$ means that the vectorizer considers unigrams and bigrams.  The RNN classifier is defined over character and word unigrams tokenizer; in addition, we further consider RNN classifiers that use and discard ``unknown'' tokens. 
The models use a common and standard preprocessing technique that removes hashtag, mentions, and URLs from tweets. Table~\ref{tab:results} summarize models' configuration.

We now briefly describe the hyperparameters selection and training strategy of the two models categories. The SGD classifier is implemented with a greed search strategy over the following TfidfVectorizer's hyperparameters: max document frequency $(0.5, 0.75, 1)$, max number of features $(1000, 5000)$, and use IDF $(True, False)$. We use the validation set to find the best configuration. RNN models implement default hyperparameters configurations: the embedding dimension is 100, and the GRU's hidden size is 256. 
The vocabulary size is set to 25K tokens for word-based cases, while 100 for character based ones; these vocabulary thresholds allows the model to learn the representation of ``unknown'' tokens. 
The training process uses Adam optimizer and BCEWithLogitsLoss as loss function. The models are trained for a maximum of 100 epochs. Note that we use a stopper mechanism that interrupts the training if a model does not improve its validation performance for 5 epochs. 

\subsection{Results and Considerations} \label{sub.res_indoor_res}
In this section, we evaluate the performance of the six models presented in Section~\ref{sub.res_indoor_res}. The first evaluation is conducted with the \textit{accuracy score} (ACC), i.e., the percentage of correct predictions. Table~\ref{tab:results} summarize the results. As expected, DNN-based models tend to outperform simple linear models; this gap can be linked with the limited vocabulary size adopted in the \textit{TfidfVectorizer} due to memory limitations. We also highlight that the usage of unknown tokens does not boost-up models' performance.

The effect of ZeW is measured with the \textit{attack success percentage} (ASP), i.e., the percentage of sentences classified as positive. Note that such a percentage also contains those samples that are misclassified in normal conditions. The evaluation uses three corpora: a set of original tweets called ``\textit{real}'', and two malicious counterparts (one per mask) named ``\textit{mask1}'' and ``\textit{mask2}'', respectively. The set ``\textit{real}'' corresponds to the negative test sentences (160K); we then discard those sentences that cannot be modified by Algorithm~\ref{Alg.1}, resulting in a final set with 75K tweets. 

Table~\ref{tab:results} presents ZeW success percentage. We can notice that the ASR is always under 40\%. This result can be explained with:
\begin{itemize}
    \item a limitation of Algorithm~\ref{sub.alg}, where the injection strategy modifies negative tokens. However, a sentence's polarity might be the effect of a sequence of tokens rather than the sum of single instances polarity.
    \item a limitation of ZeW, where the injection is limited to a strict set of operations (i.e., the insertion of a set of characters). 
\end{itemize}
Nevertheless, we can find some insights from such results:
\begin{enumerate}
    \item ZeW can affect the performance of different models that use different tokenization strategies. The combination of ZeW with state-of-the-art attacks targeting ML-models can result in dangerous effects.
    \item In general, character-based models are more resilient to ZeW. In particular, we highlight that when unknown tokens are discarded, ZeW  attack fails. 
    \item In general, models that consider unknown tokens are more vulnerable. An attacker can thus leverage this factor.
\end{enumerate}

\begin{table*}[!ht]
    \centering
    \begin{tabular}{ccc|ccc|ccc} \toprule
         \textbf{ML Model} &  \textbf{Tokenization} & \textbf{UNK} & \textbf{Train (ACC)} & \textbf{Valid (ACC)} & \textbf{Test (ACC)} & \textbf{Real (ASP)} & \textbf{Mask1 (ASP)} & \textbf{Mask2 (ASP)}\\  \midrule
         
         SGDClassifier & Char & No & 77.15 & 77.00 & 77.19 & 12.06 & 22.15 & 29.63\\
         SGDClassifier & Word & No & 73.04 & 73.00 & 73.15 & 14.94 & 20.88 & 27.12 \\ \midrule
         
         RNN & Char & No & 81.68 & 81.52 & 81.46 & 5.27 & \textbf{3.72} & \textbf{3.72} \\
         RNN & Char & Yes & 82.60 & 82.41 & 82.39 & 7.57 & 12.53 & 21.34 \\
         RNN & Word & No & 84.79 & 84.20 & 84.28 & 6.93 & 37.75 & 37.19\\
         RNN & Word & Yes & 84.93 & 84.38 & 84.41 & 6.25 & 37.29 & 36.62\\ \bottomrule
    \end{tabular}
    \caption{Overview of models' performance. The accuracy score (ACC) measure the quality of the model on the three splits. The attack success percentage (ASP) measures the misclassification percentage of a given classifier; in bold the results of models resistant to ZeW.}
    \label{tab:results}
\end{table*}

\section{Results on MLaaS}\label{sec.results} 
In this section, we show how ZeW affects the performance of different MLaaS of the leading IT companies: Amazon, Google, IBM, and Microsoft. The considered companies provide similar services, and, where possible, the results are grouped-by. 
We identified the following macro-areas.
\begin{itemize}
    \item \textit{Hate Speech Detection} (Section~\ref{sub.hs}). Tools that identify toxicity/hate speech in comments.
    \item \textit{Insights Extractors} (Section~\ref{sub.oth}). Tools that extract insightful information from the text (e.g., tones, personalities). 
    \item \textit{Sentiment Analyzers} (Section~\ref{sub.sent}). Tools that measure sentence polarization. 
    \item \textit{Translators} (Section~\ref{sub.tr}). Tools that translate sentences from a source language to a target one. 
\end{itemize}

In this work, we do not compare the performance of ZeW with state-of-the-art since their focus is to exploit ML algorithms vulnerabilities, while we aim at the disruption of the indexing stage.
Since our attack model is free of all of the restrictions in the number of modifications, an attacker can combine ZeW with attacks targeting ML algorithms.


\subsection{Dataset \& Evaluation on VaderSentiment}\label{sub.vadRes}
For the experiments, we use the hateful sentences available in~\cite{ICWSM1715665}, a well-known dataset of the task. This dataset contains three distinct classes: ``hateful'', ``offensive but not hateful'', and ``neither'' (nor hateful neither offensive). The dataset includes 1430 hateful sentences. We call now on the set of hateful sentences \textit{Real}. Sentences that do not contain any negative word (detected) are discarded from \textit{Real}. We then applied the injection algorithm with the two possible masks, generating two sets called, respectively, \textit{Mask1} and \textit{Mask2}. The final corpora contain 1094 samples each. All of the analyses and tests in different MLaaS use these corpora.

\begin{figure}[!htb]
    \centering
    \includegraphics[width=.8\linewidth]{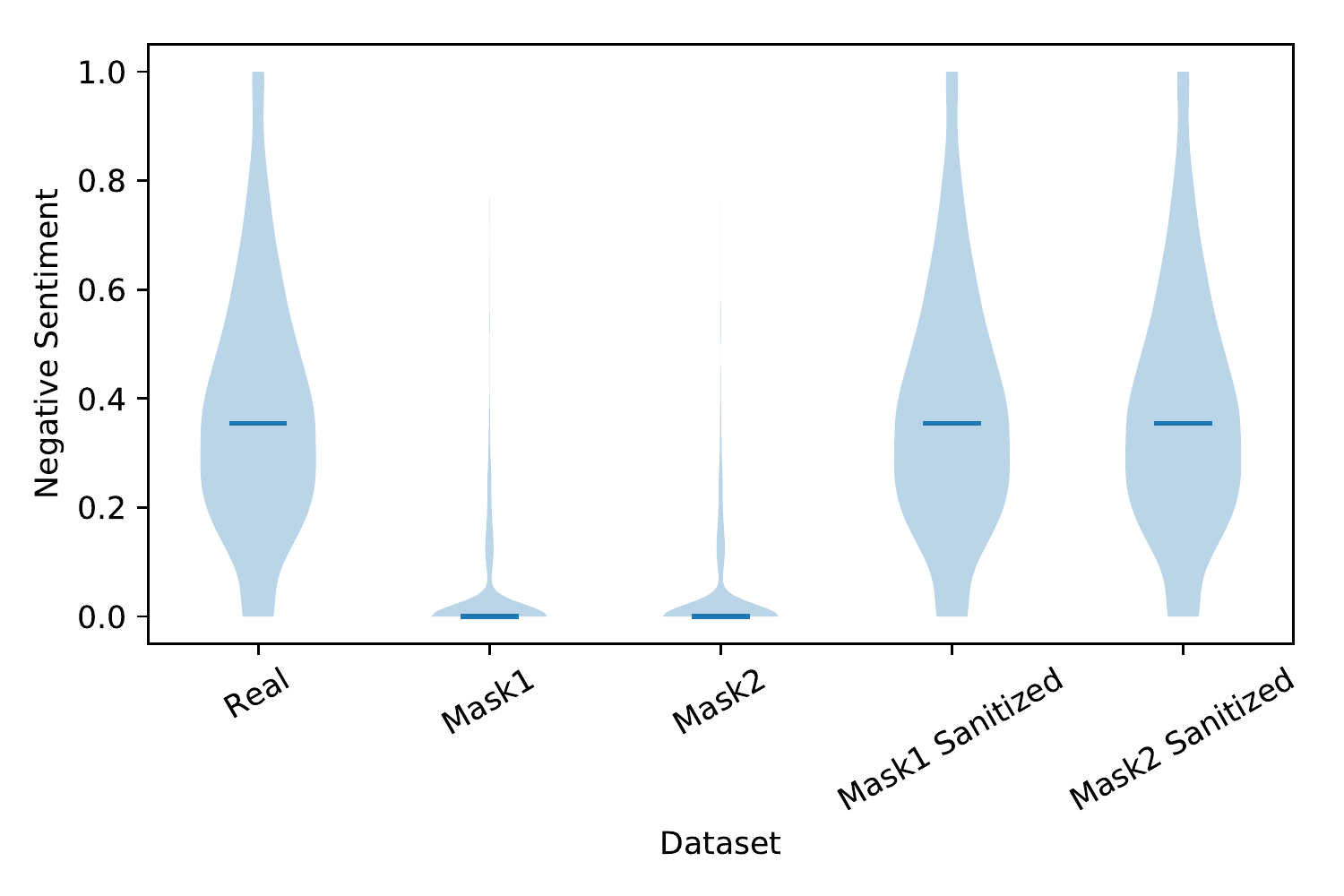}
    \caption{Negative sentiment densities of different corpus measured by VaderSentiment, where +1.0 is extremely negative and 0.0 is absence of negativity. The violin plot shows the distributions of the corpora's negative scores; the blue line represents the median value of the distribution. The service is vulnerable if the distributions under attack are not equal to the distribution of \textit{Real}.}
    \label{fig:vader}
\end{figure}

We first analyze the impact of ZeW on VaderSentiment. As shown in Figure~\ref{fig:vader}, both injection strategies (\textit{Mask1} and \textit{Mask2}) entirely cancel the perceived negativity. The median values of negativity scores are 0.35 (\textit{Real}), and 0.0 for both \textit{Mask1} and \textit{Mask2}. ZeW is effective in both modalities against VaderSentiment. The injection of only one character per negative word is enough to disrupt this service. This might be a relevant problem since this tool is widely used in the scientific community.

In this section, we further show the effectiveness of our defense strategies proposed in Section~\ref{sub.counter}, where the sanitization technique discards the malicious character from any given sentence. Figure~\ref{fig:vader} shows that the sanitized sentences have the same distribution of the original and unpoisoned corpus. Given the simplicity and the effectiveness of the proposed countermeasure, we decide to do not report similar results in the rest of  Section~\ref{sec.results}.

\subsection{Hate Speech Detection}\label{sub.hs}
We start our analysis with hate-speech detection, the set of tools closer to our case study. These tools aim is to identify and detect the toxicity of sentences. The goal of an adversary is to write hateful sentences without being detected. 
This scenario is likely on Social Networks (e.g., Facebook) that uses account suspension or ban when users post inappropriate content.
We analyzed the following services.
\begin{itemize}
    \item Google Perspective\footnote{\url{https://www.perspectiveapi.com}}. Perspective is part of the Conversation AI project, which aim is to improve the quality of online conversations with the supervision of ML. The tool identifies several aspects of online conversations that might be inappropriate, such as toxicity, profanity, and flirtation. In this experiment, we focus on toxicity manipulation, defined as disrespectful comments.
    \item Microsoft Content Moderator\footnote{\url{https://azure.microsoft.com/en-us/services/cognitive-services/content-moderator}}. This ensemble of ML-tools aim to identify potentially offensive content in different type of media, such as text, images, and videos. Regarding the text-domain, the model identifies three categories of malicious content: sexually explicit content, sexually suggestive, and offensive. In this analysis, we focus on the latter. The tool offers the ``autocorrect'' option, which corrects grammatical mistakes before analyzing the contents. In our experiment, this parameter is set to TRUE.   
\end{itemize}

\begin{figure*}[!ht] 
    \centering
    \subfloat[Google Perspective.]{%
        \includegraphics[width=0.4\linewidth]{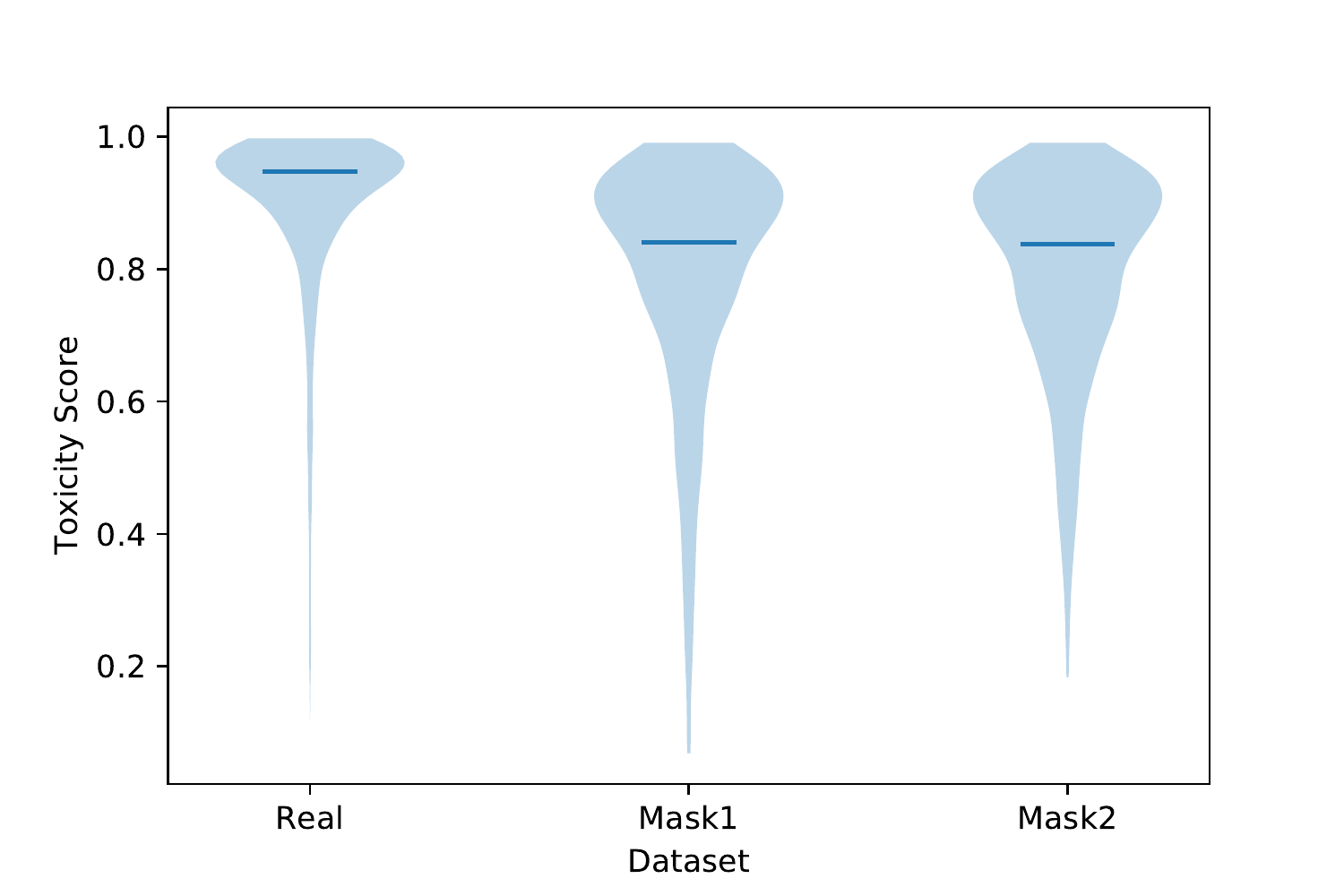}%
        \label{fig:persp}%
        }%
    \hfill%
    \subfloat[Microsoft Moderator.]{%
        \includegraphics[width=0.4\linewidth]{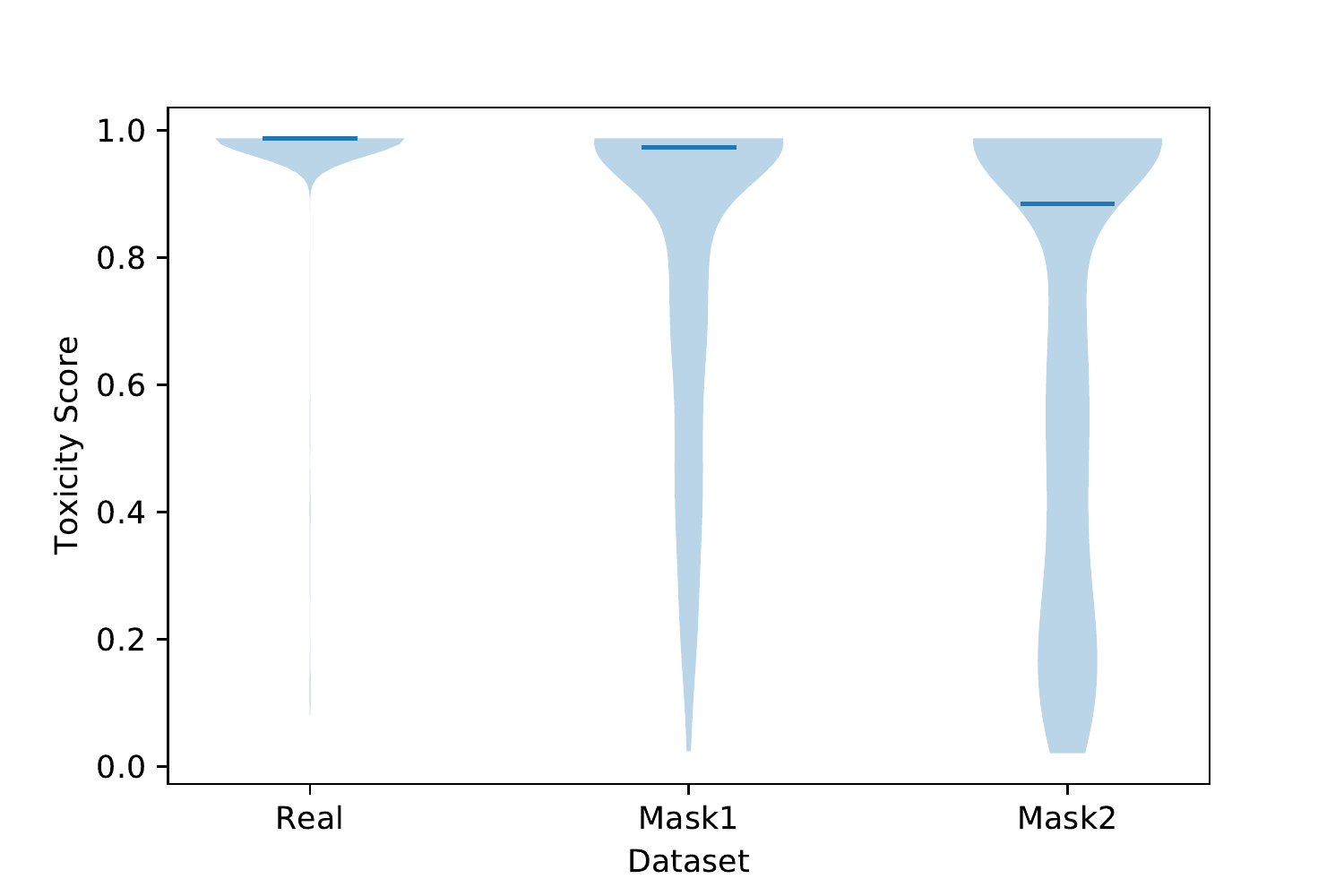}%
        \label{fig:mic_mod}%
        }%
    \caption{Toxicity score densities of different corpora measured by Google Perspective (left), and Microsoft Moderator (right), where +1.0 is high confidence of being classified as toxic. The violin plot shows the distributions of the corpora’s toxicity scores; the blue line represents the median value of the distribution. A service is vulnerable if the distributions under attack are not equal to the distribution of \textit{Real}.}
    \label{fig:toxicity}
\end{figure*}

Figure~\ref{fig:toxicity} shows the effect of ZeW on the toxicity detectors. Both services are highly resistant to the attack.
On Google Perspective the median confidence level of the detector is 0.95 on (\textit{Real}), 0.84 on \textit{Mask1}, and 0.83 on (\textit{Mask2}). Similarly, on Microsoft Moderator the median is 0.99 on (\textit{Real}), 0.97 on \textit{Mask1}, and 0.86 on (\textit{Mask2}). 

The impact of the attack is not strong, and the model seems resistant. On the other hand, in Google Perspective the insertion of only one zero-width character per negative word appears sufficient to damage the model's confidence level. Similarly, the Microsoft tool can be affected by \textit{Mask2}. We also need to highlight that the purpose of these tools is to detect high toxicity levels rather than detect negativity on sentences; thus, the algorithm described in Figure~\ref{Alg.1} might not be effective.  
The combination of ZeW with other state-of-the-adversarial techniques could seriously damage this service. We can state that both models are vulnerable to this attack.

\subsection{Insights Extractors}\label{sub.oth}
Online Social Networks (OSNs) such as Facebook and Twitter are places where billions of users share their experiences, ideas, feelings, and opinions. These platforms are perfect for analyzing social behaviors and interactions. Several studies are conducted, from sentiment analysis and opinion mining~\cite{pak-paroubek-2010-twitter-corpus, 10.1007/978-3-642-35176-1_32}, to the prediction of when a security vulnerability will be exploited~\cite{10.1145/3292500.3330742}. 
IBM offers two services that are helpful to analyze OSNs data.
A possible attacker's goal is to hide his/her own personality.
\begin{itemize}
    \item IBM Watson Tone Analyzer\footnote{\url{https://www.ibm.com/cloud/watson-tone-analyzer}}. The tool detects and extracts emotional and language tones in a written text.
    \item IBM Watson Personal Insight\footnote{\url{https://www.ibm.com/watson/services/personality-insights}}. The tool predicts the personality of a target user. For example, this tool allows us to analyze the tweets-history of a target Twitter account. 
\end{itemize}

\begin{figure}
    \centering
    \includegraphics[width=.8\linewidth]{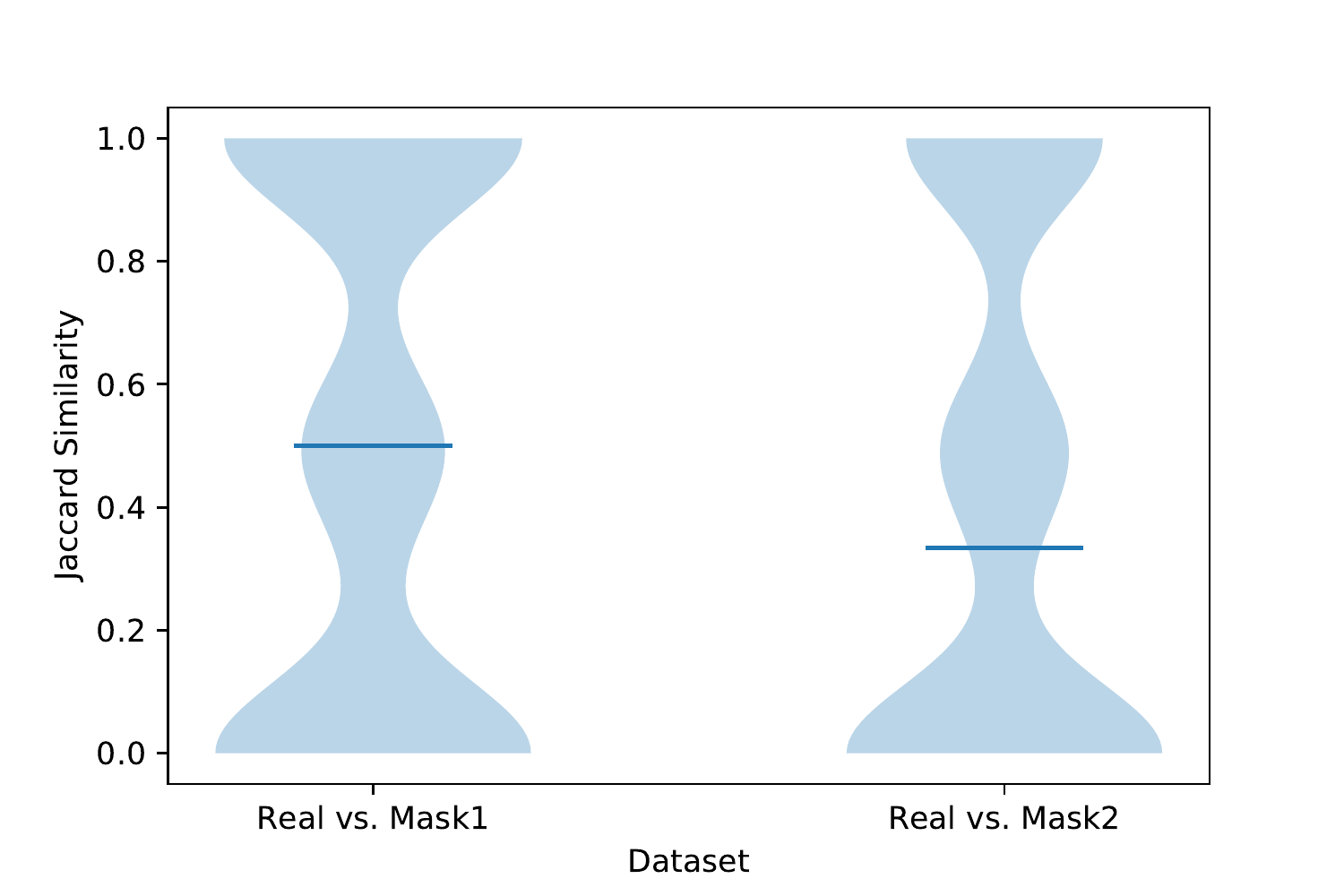}%
    \caption{The similarity distributions between \textit{Real} vs. \textit{Mask1} and \textit{Real} vs. \textit{Mask2} of Watson Tone Analyzer, where +1,0 is an exact match between two sets. The violin plot shows the distributions of the corpora’s Jaccard similarities; the blue line represents the median value of the distribution. The service is vulnerable if the distributions under attack are not close to one.}
    \label{fig:ta}
\end{figure}

IBM Watson Tone Analyzer returns a list of emotions (strings) detected in a given sentence. Here, a possible adversary's goal is to hide/manipulate emotions from his text. To understand the efficacy of ZeW, we measured the similarity between the sets of emotions of the unpoisoned sentences and their poisoned counterparts. In particular, given a sentence $x$, its adversarial counterpart $x'$, and a Tone Extractor function $f$, we obtain the sets $A=f(x)$ and $B=f(x')$. The similarity between $A$ and $B$ is given by the \textit{Jaccard Similarity}, defined as follows~\cite{niwattanakul2013using}.
\begin{equation}
    J(A, B) = \frac{dim(A\cap B)}{dim(A \cup B)},
\end{equation}
where $dim$ returns the number of items in the set.
The performance is measured by comparing the Jaccard similarities of \textit{Real} vs. \textit{Mask1} and \textit{Real} vs. \textit{Mask2}. 
Ideally, the API is resistant if the Jaccard Similarity is equal to +1.0 (two identical sets). 
In Figure~\ref{fig:ta}, we can notice a different trend. The median values are 0.5 and 0.33 for \textit{Mask1} and \textit{Mask2}, respectively. 
A good portion of sentences (40\%) are not affected; a possible explanation is that the negative words of those sentences are not essential to extract the emotion.
Note that from this analysis we discard those sentences without any ``tone'' detected by the tool (322 sentences discarded). 

\begin{figure}
    \centering
    \includegraphics[width=1.\linewidth]{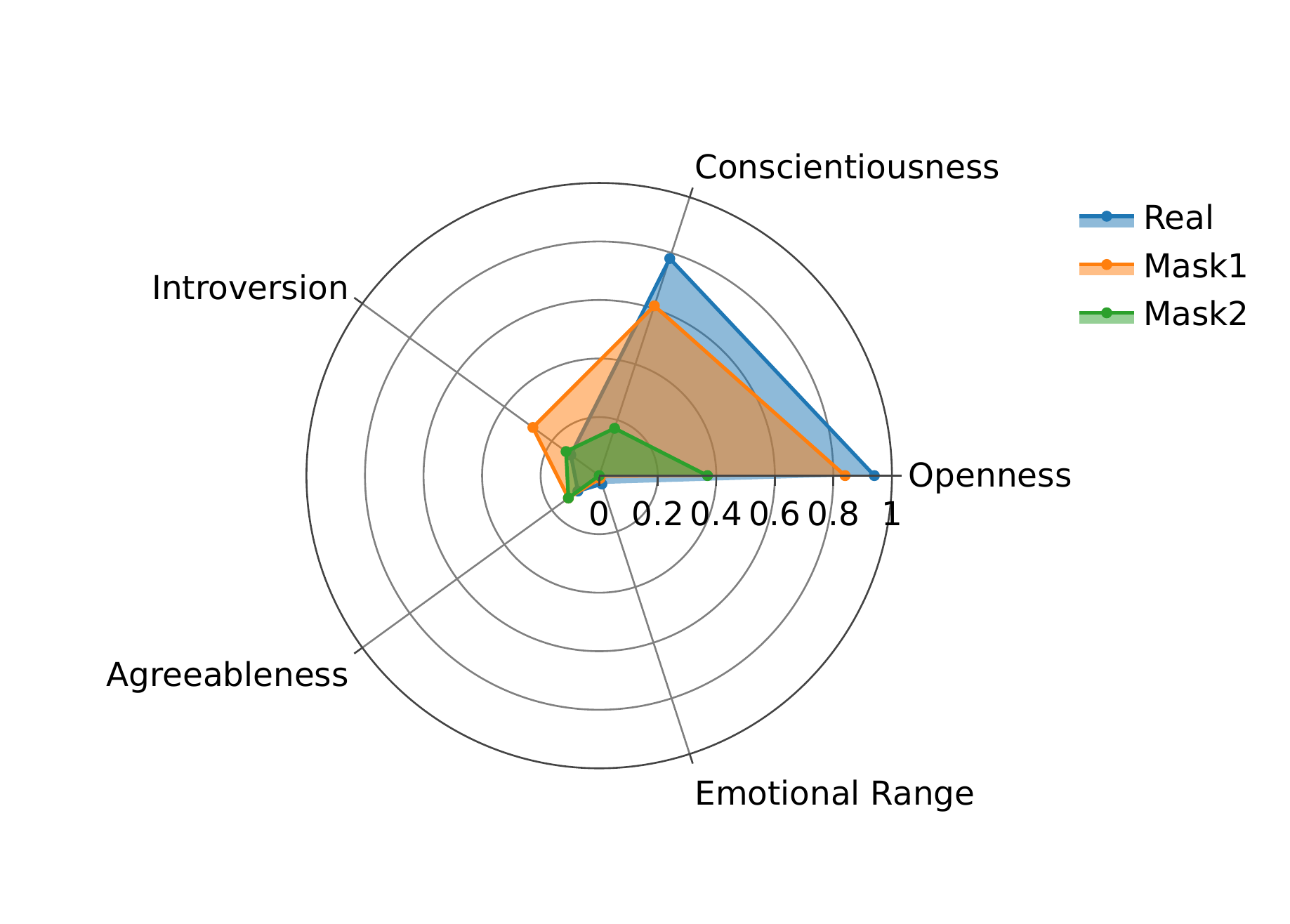}
    \caption{ Watson Personal Insight detects three distinct personalities (\textit{Real}, \textit{Mask1}, and \textit{Mask2}). The service is vulnerable if at least one of the five dimensions changes.}
    \label{fig:personalities}
\end{figure}

On IBM Watson Personal Insight, the adversary's goal is to hide/manipulate his personality. In our test, we extract the personalities from the three corpora. In this experiment, the analysis is at a corpus-level rather than a sentence-level, i.e., we obtain one personality for each corpus. In Figure~\ref{fig:personalities}, we can notice that \textit{Real} and \textit{Mask1} differ in terms of ``Openness'' and ``Conscientiousness'', while \textit{Mask2} seems to push all of the dimensions close to zero. In conclusion, both services of IBM are severely vulnerable to ZeW.

\subsection{Sentiment Analyzers}\label{sub.sent}
Sentiment analysis is one of the most popular topics in NLP~\cite{7724305, liu2012sentiment, agarwal2011sentiment, maas2011learning} and can be used for several purposes, such as understand the opinions of restaurants, movies, or products. This importance is reflected by the fact that all companies implement this service: Amazon Comprehend\footnote{\url{https://aws.amazon.com/comprehend}.}, Google Cloud Natural Language\footnote{\url{https://cloud.google.com/natural-language}.}, IBM Watson Natural Language Understanding\footnote{\url{https://www.ibm.com/cloud/watson-natural-language-understanding}.}, and Microsoft Text Analytics\footnote{\url{https://azure.microsoft.com/en-us/services/cognitive-services/text-analytics}.}. 

In the hate speech scenario, as shown in Figure~\ref{fig:vader}, the sentences are likely to be perceived as negative. A possible adversary's goal is to minimize the detected negativity; this attack can be seen as a \textit{transferable attack}~\cite{papernot2016transferability} since our malicious sentences are first tested on a sentiment analyzer (i.e., VaderSentiment). 

\begin{figure*}[!t] 
    \centering
    \subfloat[Amazon Comprehend.]{%
        \includegraphics[width=0.4\linewidth]{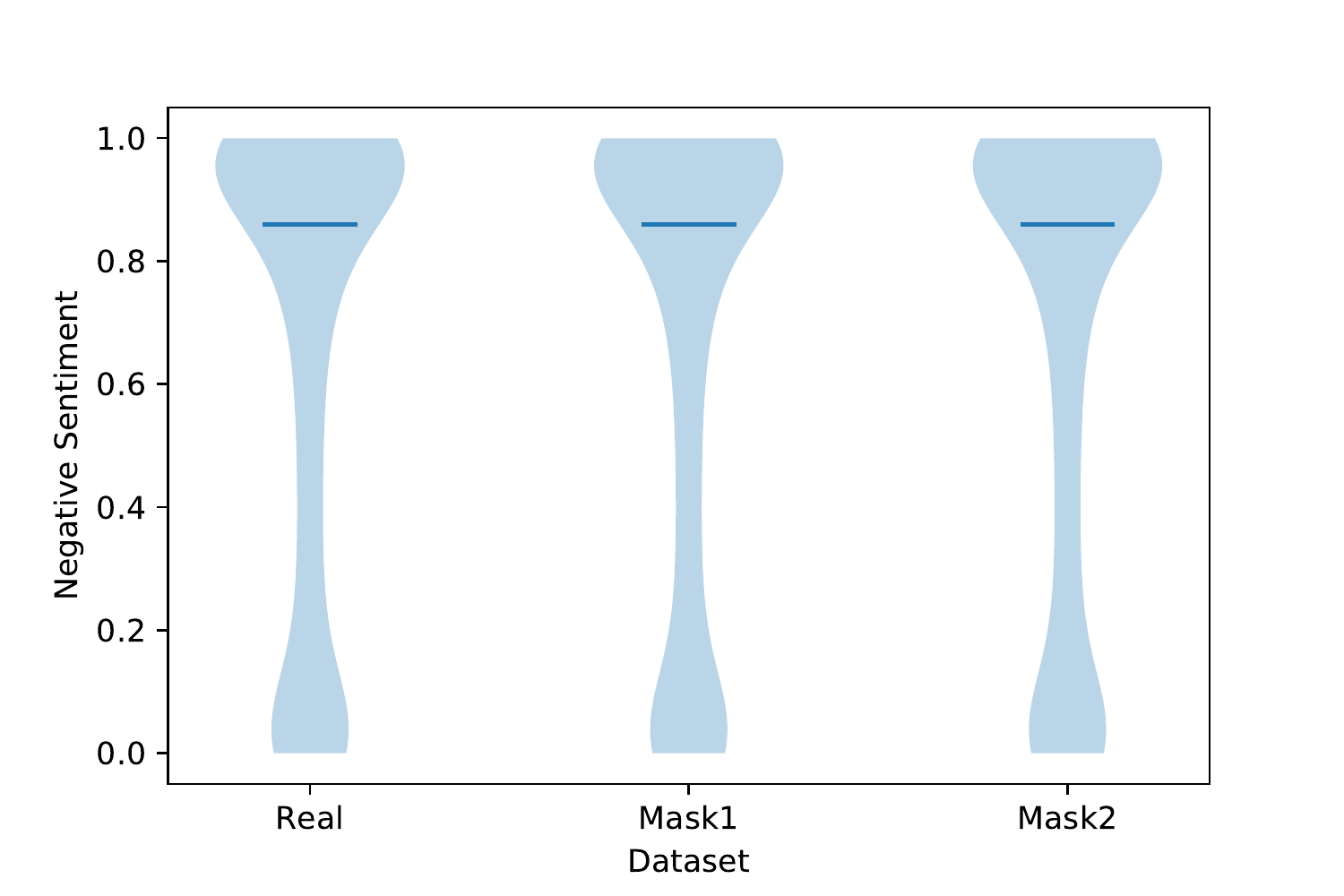}%
        \label{fig:amazon_comprehend}%
        }%
    \hfill%
    \subfloat[Google Cloud Natural Language.]{%
        \includegraphics[width=0.4\linewidth]{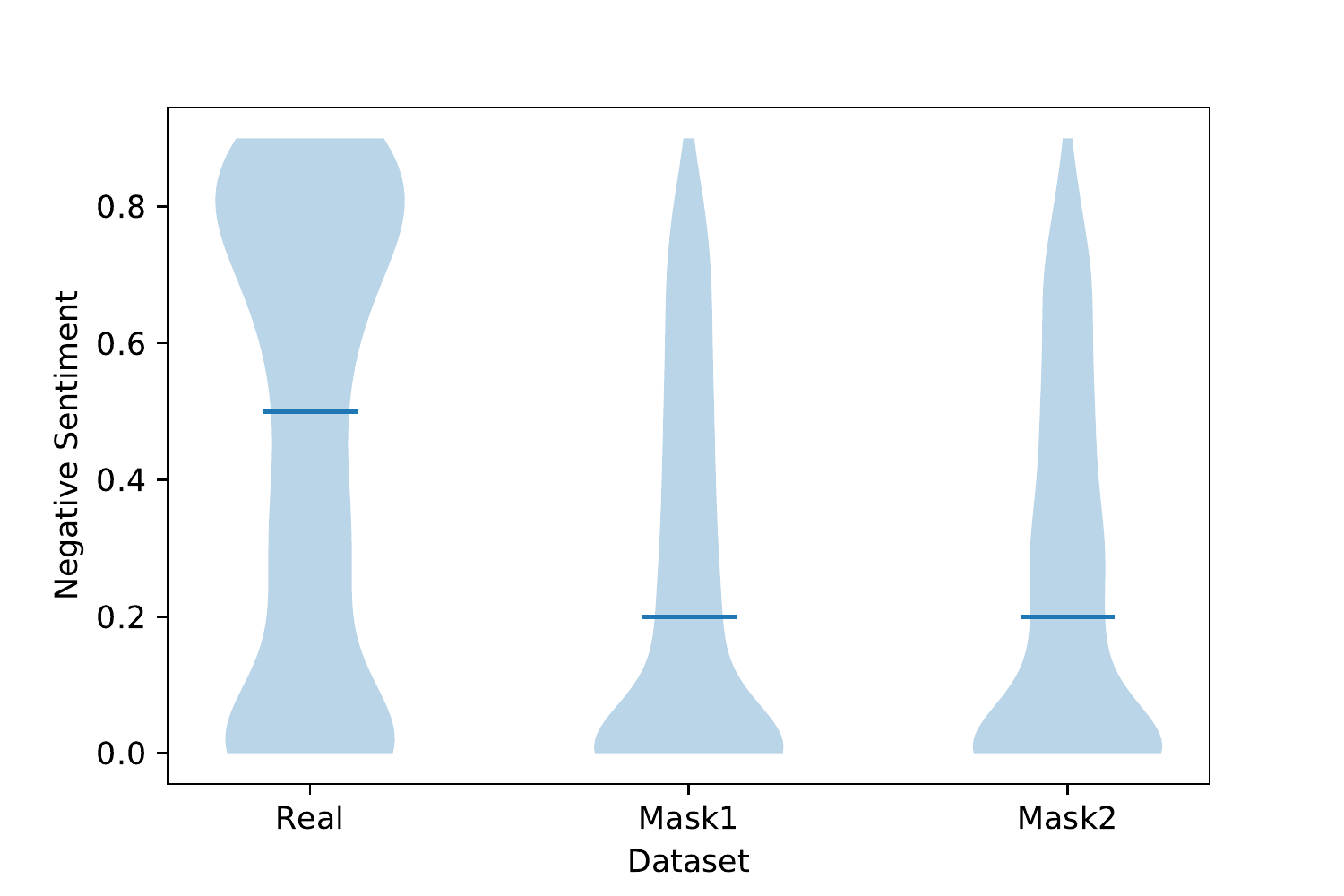}%
        \label{fig:google_cnl}%
        }
    \newline
    \subfloat[IBM Watson Natural Language Understanding.]{%
        \includegraphics[width=0.4\linewidth]{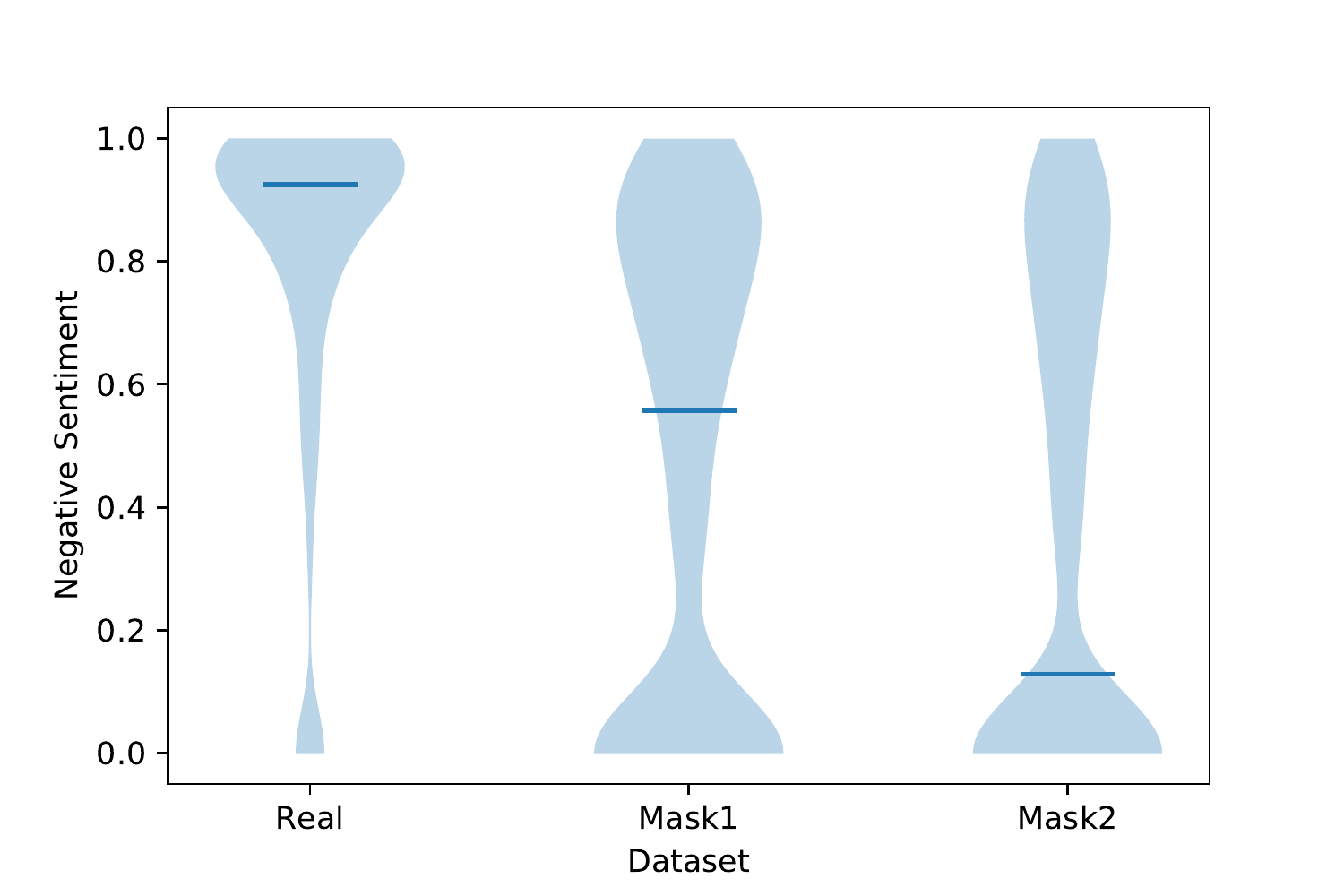}%
        \label{fig:watson_nlu}%
        }%
    \hfill%
    \subfloat[Microsoft Text Analytics.]{%
        \includegraphics[width=0.4\linewidth]{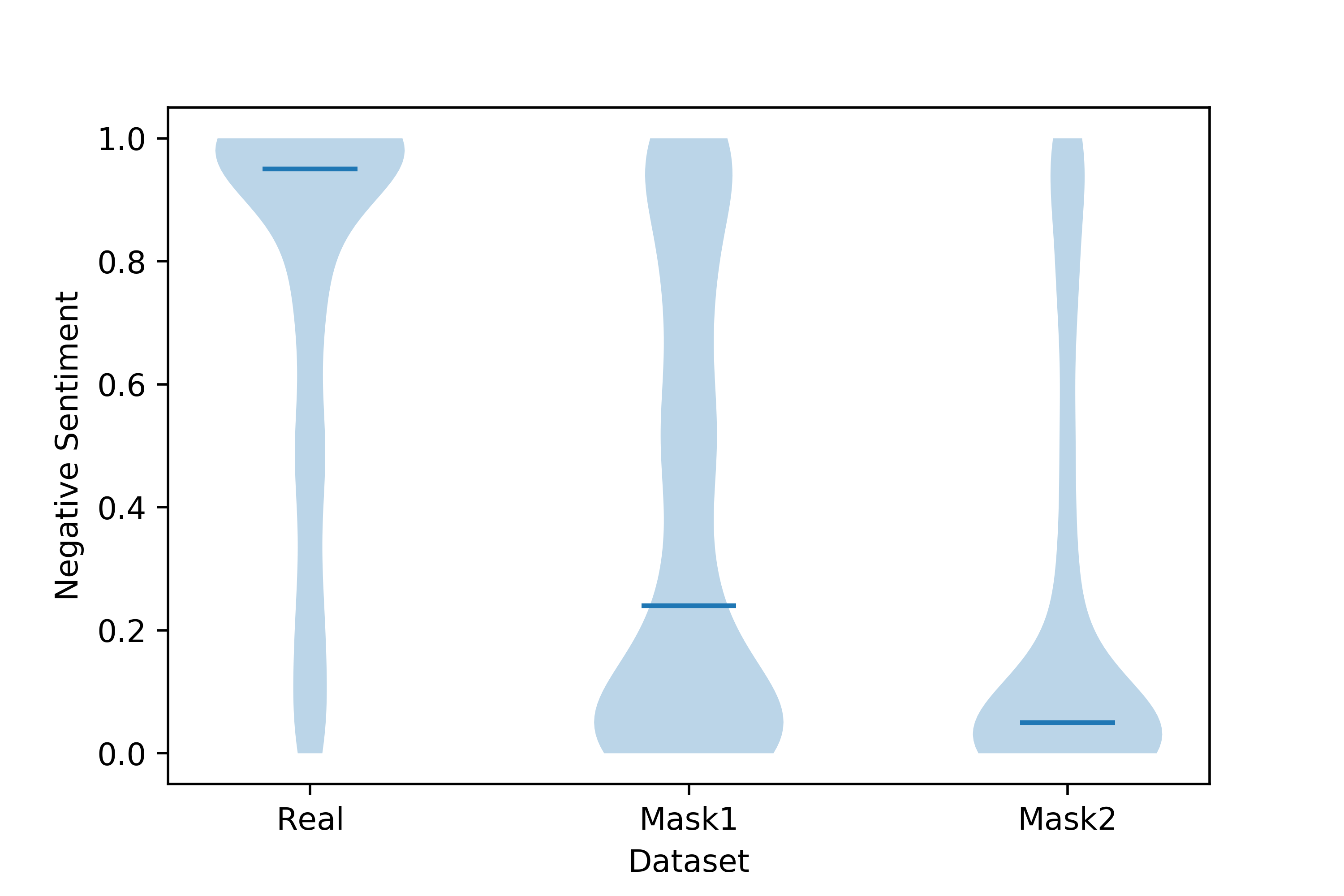}%
        \label{fig:microsoft_ta}%
    }
    \caption{Effect of Zero-Width Space Attack on different sentiment extractor services. The violin plot shows the distributions of the corpora’s negative scores; the blue line represents the median value of the distribution. A service is vulnerable if the distributions under attack are not equal to the distribution of \textit{Real}.}
    \label{fig:sent}
\end{figure*}

Figure~\ref{fig:sent} shows the effectiveness of ZeW on 3 services out of 4. In particular, Amazon Comprehend is resistant to both modalities of injection, where the median value is constant (0.86). Google Cloud Natural Language shows a similar vulnerability pattern for both masks, with an equal median value that moves from 0.5 to 0.2. 
In this service, the addition of one character per negative word is sufficient to disrupt it. We conclude with the services provided by IBM and Microsoft, where we see a common decreasing pattern of the median values, which move from 0.92 / 0.95 on \textit{Real}, to 0.56 / 0.24 for \textit{Mask1}, and to 0.13 / 0.05 for \textit{Mask2}. 
We can state that three out of four services are severely vulnerable to ZeW, while only one show resistance. 

\subsection{Translators}\label{sub.tr}
We conclude the results section with another well-known NLP task: the language translation. All four companies implement this service: Amazon Translate\footnote{https://aws.amazon.com/translate.}, Google Translation\footnote{https://cloud.google.com/translate?hl=en.}, IBM Watson Language Translator\footnote{https://www.ibm.com/cloud/watson-language-translator.}, and Microsoft Translator\footnote{https://azure.microsoft.com/en-us/services/cognitive-services/translator.}. 

In the hate speech scenario, we can imagine that the adversary writes a hateful message in an unknown language for the victim. The victim uses translators to understand the meaning of the message. For example, human moderators could use automatic translators to understand if comments written in foreign languages are hateful. Another example is browsers like Chrome that automatically translates web content.

\begin{figure*}[!t] 
    \centering
    \subfloat[Amazon Translate.]{%
        \includegraphics[width=0.4\linewidth]{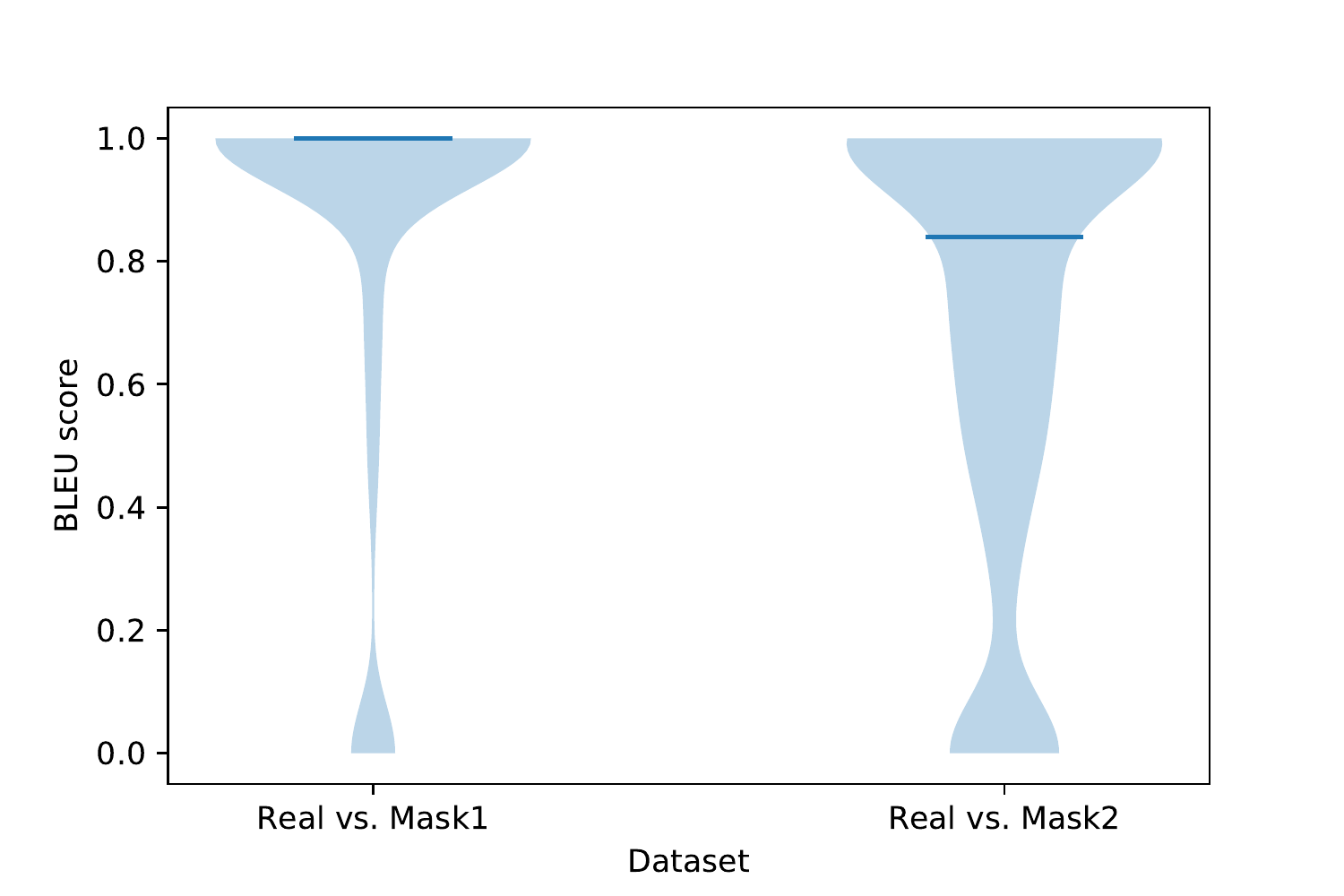}%
        \label{fig:amazon_tr}%
        }%
    \hfill%
    \subfloat[Google Translate.]{%
        \includegraphics[width=0.4\linewidth]{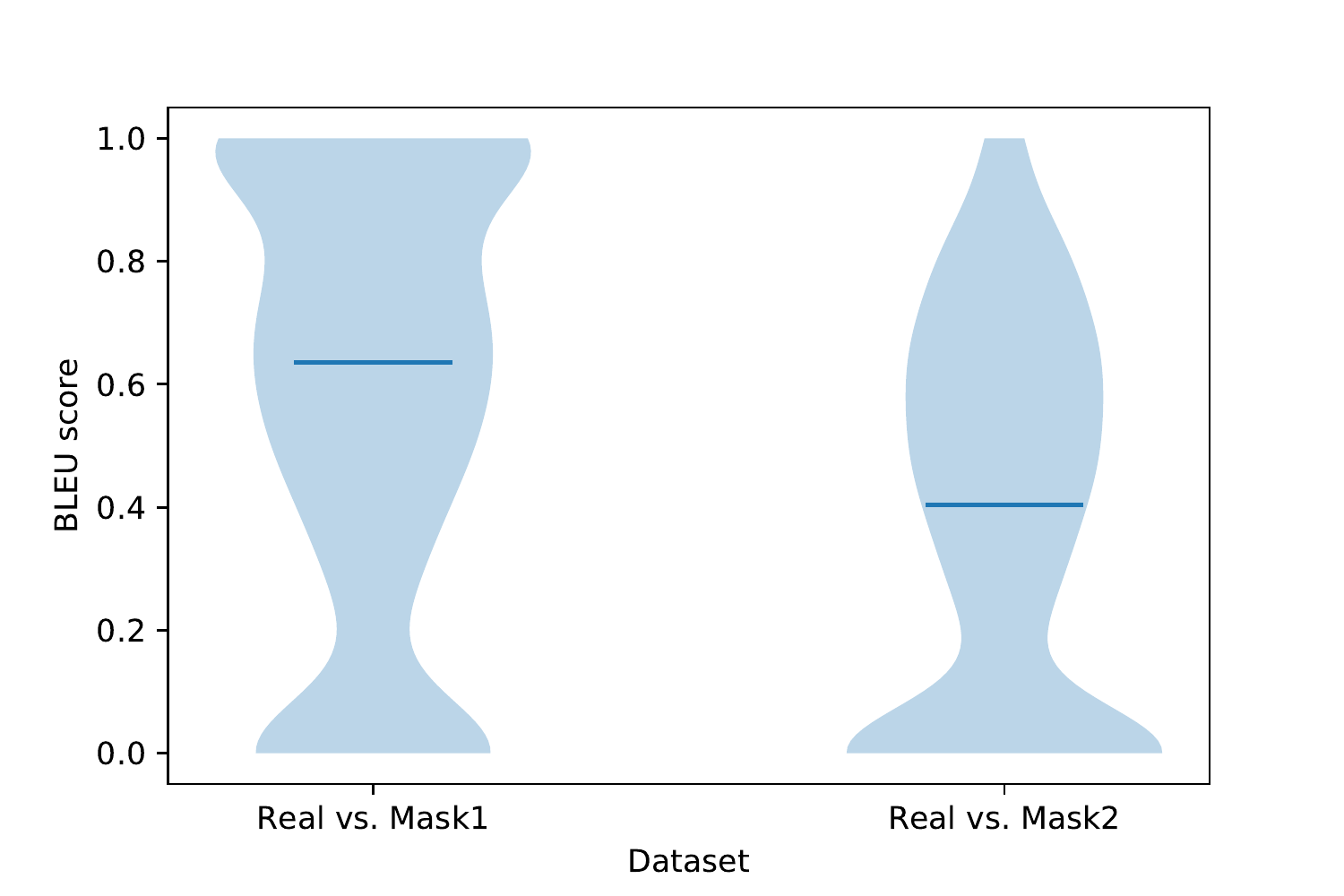}%
        \label{fig:google_tr}%
        }
    \newline
    \subfloat[IBM Watson Natural Language Translator.]{%
        \includegraphics[width=0.4\linewidth]{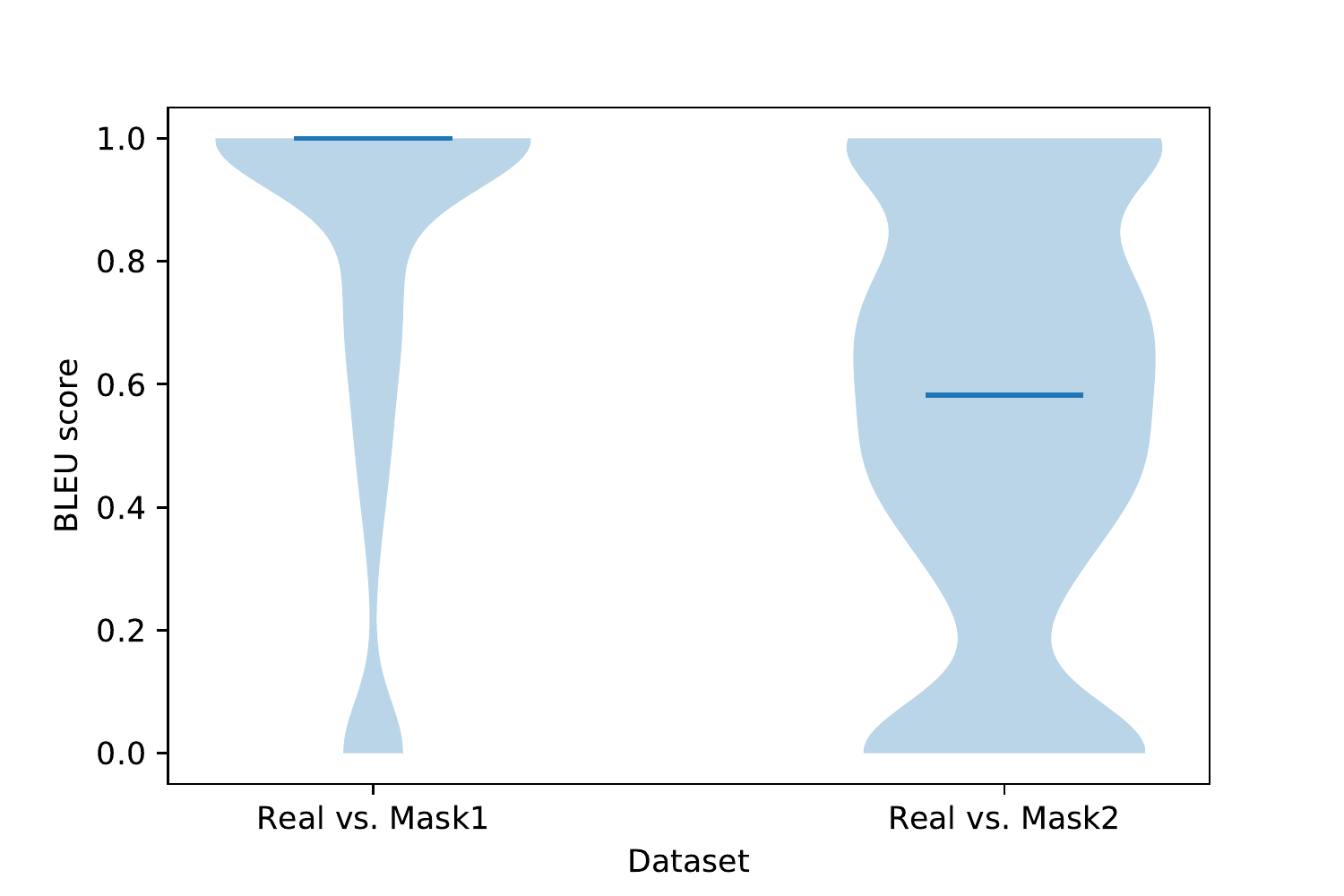}%
        \label{fig:watson_tr}%
        }%
    \hfill%
    \subfloat[Microsoft Translator.]{%
        \includegraphics[width=0.4\linewidth]{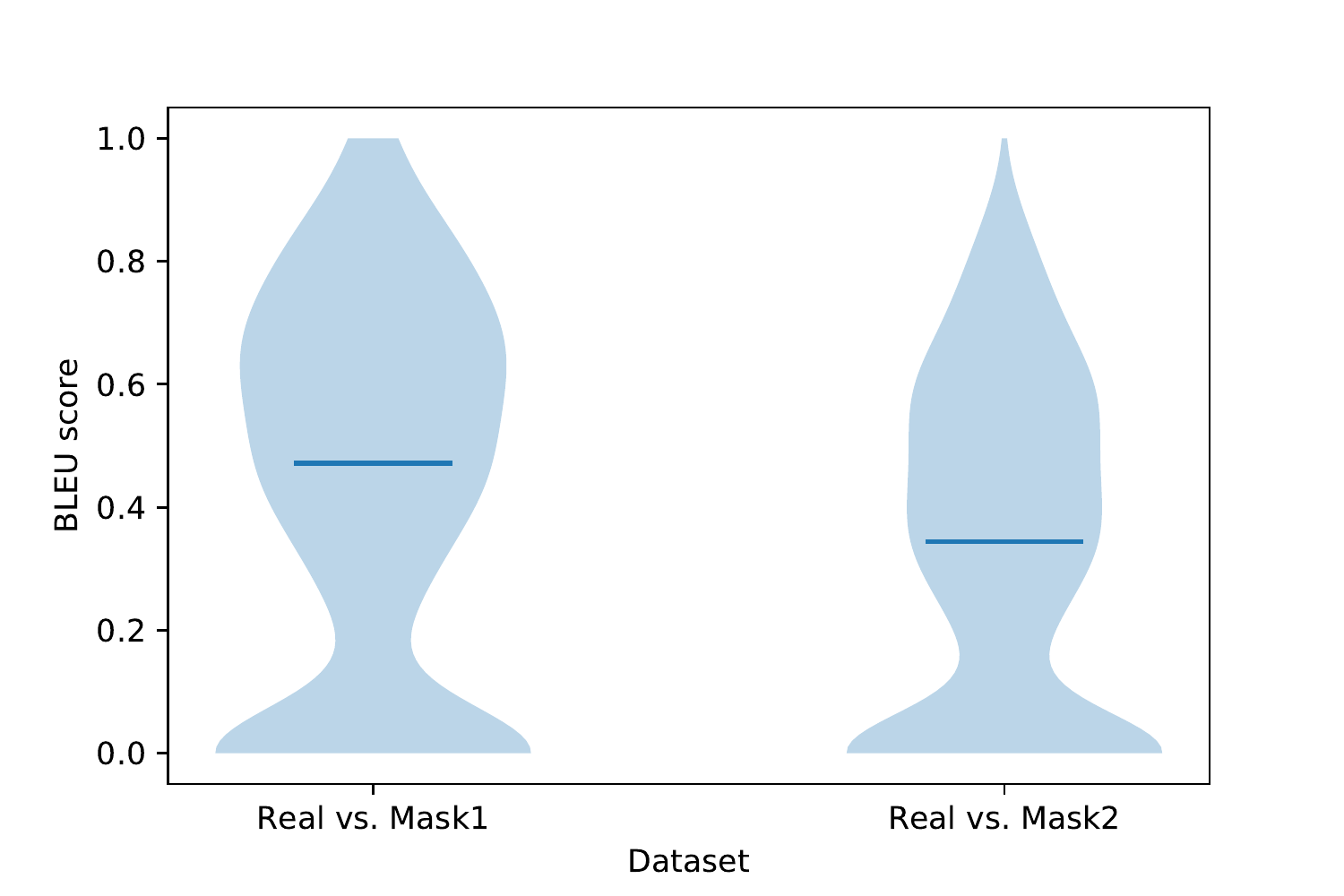}%
        \label{fig:microsoft_tr}%
    }
    \caption{Effect of Zero-Width Space Attack on different translator services. The violin plot shows the distributions of the corpora’s BLEU scores; the blue line represents the median value of the distribution. A service is vulnerable if the distributions under attack are not close to one.}
    \label{fig:transl}
\end{figure*}

An example of this scenario is shown in Figure~\ref{fig:g_translate}, where the malicious sentence ``I wanna kill you'' is translated as ``I love you'' by Google Translate\footnote{https://translate.google.com/.}. Note that since the target model is unknown, we do not have any control over the target output. We highlight here that the aim of the attacker is to degrade the general performance of the target model rather than control the translation process. 
ZeW is evaluated on the translation task English-Italian. 
To understand the impact, we measure the similarity between the translations given by the unpoisoned sentence and its malicious counterpart. The difference is measured with the Bilingual Evaluation Understudy Score (BLEU score), with its 4-gram cumulative implementation. Formally, given a sentence $x$, its malicious counterpart $x'$, and a translation function $f$, the similarity is defined as
\begin{equation}
    similarity = BLEU4(f(x), f(x')).
\end{equation}
Ideally, a service is not affected if the translations of the original sentence and its malicious version are the same, resulting in BLEU score equal to +1.0 (perfect match). 
In Figure~\ref{fig:transl} we can see that all of the services are vulnerable to the attack. Amazon seems resistant to \textit{Mask1}, with a median value equal to 1.0, while vulnerable to \textit{Mask2}, with the median equal to 0.83. Similarly, IBM is resistant to \textit{Mask1} and vulnerable to \textit{Mask2}: the median value is 1.0 for \textit{Mask1}, and 0.58 for \textit{Mask2}.  
Google and Microsoft show vulnerabilities in both injection strategies, where the median values move from 0.63 / 0.47 in \textit{Mask1}, to 0.40 / 0.34 in \textit{Mask2}.

All of the models show more difficulties in handling \textit{Mask2}. These tools show different vulnerability patterns compared to the sentiment analysis tasks. The possible explanation is the nature of translators: Seq2Seq models (i.e., autoencoders). Seq2Seq models likely use different placeholders to deal with OOV tokens, as introduced in Section~\ref{sub.textpipeline}.
We can state that all of the services are vulnerable to ZeW: three strongly vulnerable, and only one weakly (Amazon).


\subsection{Considerations}\label{sub.res_cons}
In this section, we analyzed how different MLaaS behave under the ZeW attack. We can notice different trends among types of services (e.g., sentiment analyzers) and the same companies (e.g., Microsoft). We now try to understand why these models behave differently. 

ZeW seems to fail on \textit{hate speech detectors}. This result suggests that both services use character-based tokenizers, which is a reasonable assumption since such services should deal with noisy text (e.g., grammatical errors, misspelling) gathered from blogs, forums, and social networks. Moreover, such services are resistant to the injected noise (unknown tokens); a possible explanation is that these services deal with unrecognized words (e.g., discard).
Amazon services show similar performance.

In general, IBM MLaaS are vulnerable to ZeW attack. Similar trends are shared among different services (e.g., Watson Personal Insight, sentiment extractor), where the attack is more effective when we inject more ZeW characters. These trends are similar to the RNN char-based with UNK performance, as shown in Table~\ref{tab:results}.

Finally, on translators, we find two patterns: i) resistant only to \textit{mask1} (i.e., Amazon, IBM), and vulnerable to both injection levels (i.e., Google, Microsoft). Since \textit{mask2} has a stronger impact, the four models might be character-based. However, it is unclear why there is such discrepancy, where two out of four models are resistant to \textit{mask1} ZeW attack. More in-depth investigations should be conducted on neural machine translators architecture.

%% file: Sections/Related_Works.tex
\section{Related Work}\label{sec.rel_work}
In the state-of-the-art we can find several adversarial attacks targeting the machine learning algorithms of MLaaS. 
We now briefly summarize attacks on the MLaaS considered in Section~\ref{sec.results}. 

\paragraph{\textit{Hate Speech Detectors}}Several scientific discussions use Google Perspective as a case study of their hate speech evasion techniques. For example, in~\cite{hosseini2017deceiving, 10.1145/3270101.3270103} the authors show the ability to manipulate Perspective by adding small mistakes to the sentences (e.g., typos, leet speech, word addition, word removal). In~\cite{9071126}, the authors proposed evasion techniques based on acoustic and visual similarities, with an evasion power equal to 33\% and 72.5\%.  

\paragraph{\textit{Sentiment Analyzers}}In~\cite{gong2018adversarial}, to manipulate sentiment tools, the authors applied techniques from computer visions, i.e., Fast Gradient Sign Attack~\cite{goodfellow2014explaining}. In DeepFool~\cite{moosavi2016deepfool}, the authors manipulate the sentiment analysis of a CNN model. This algorithm uses Word Mover's Distance (WMD)~\cite{kusner2015word} to find suitable words whose embeddings allow to influence the target classifier. Similarly, in~\cite{Alzantot_2018}, the authors propose a word replacement algorithm based on semantic similarities. In~\cite{DBLP:conf/ndss/LiJDLW19}, authors describe TextBugger, a black-box framework that achieves high evasion success rate on different Machine-Learning-as-a-Services.   
\paragraph{\textit{Machine Translators}}Cheng et al. propose AdvGen, a gradient-based method for attacking Neural Machine Translation (NMT) models~\cite{Cheng_2019}. In~\cite{Cheng_2020}, the authors propose two techniques to evade Seq2Seq models (e.g., translators) using ad-hoc loss functions: \textit{non-overlapping attack} and \textit{keyword attack}. For the first, the goal is to generate completely novel adversarial sentences, while for the latter, the malicious translation contains target keywords. For the interested reader, we suggest finding more details on adversarial machine learning in Seq2Seq models in~\cite{Cheng_2020}.

%% file: Sections/Limitations.tex
\section{Limitations}\label{sec.limitations}
In this section we briefly discuss the limitations of ZeW attack and the proposed coutermeasure. 

\paragraph{\textit{Attack}}
The results presented in Section~\ref{sec.results} show how different commercial services can be affected by the proposed attack ZeW. However, the efficacy of ZeW is strictly related to services implementation choices. For example, as shown in Section~\ref{sec.res_indoor}, char-based models are more resilient compared to word-based ones. Moreover, when the model discards unrecognized characters, the attack is completely unsuccessful.
Another major drawback is the limited control over malicious samples and, as a consequence, over the effect of the attack. If we consider language translators, an attacker can affect the translation, but he/she has no control over the output. For example, in the attack reported in Figure~\ref{fig:g_translate} we did not target that particular translation. 
Similarly, on the classification task, an attacker can only reduce the likelihood of a sentence being in a specific class (e.g., in this work we reduce sentences' negativity) and not let the sample be classified as a target class.

\paragraph{\textit{Defense}}
In Section~\ref{sub.counter}, we present a simple yet effective countermeasure, consisting on the removal (sanitization) of zero-width characters from any given sentence. This choice is possible since normal English sentences should not contains such characters. Moreover, to understand if a ZeW attack is occurring, models owners can feed their applications with both original and sanitized sentences and look for results discrepancies. 
The proposed sanitization technique is however applicable only for ZeW attacks, resulting in a patch rather than a general solution. 
A popular countermeasure adopted in the state-of-the-art is the \textit{adversarial training}, where, for example, the defender augments the training data with examples of adversarial samples to make the model more robust~\cite{goodfellow2014explaining}. Even though the adversarial training showed promising results, we believe that a strong and simple countermeasure consists of limiting applications' character vocabulary. 
We recall that our attack uses characters that are not normally present in the written language, and thus a simple input control can raise alerts whenever unlikely characters are identified.
Finally, as reported in Table~\ref{tab:results}, character-based models present an intrinsic resiliency to ZeW attack; future commercial implementation should consider this aspect. 

%% file: Sections/Conclusions.tex
\section{Conclusions}\label{sec.end}
The migration of machine learning applications from research to commercial and industrial purposes increases the necessity of finding security mechanisms that guarantee the correct usage of them.
In this work, we present a novel injection algorithm: Zero-Width attack (ZeW). This attack injects non-printable UNICODE characters on malicious sentences, with a potential disruption of the indexing stage of the ML application pipeline, while maintaining the full-readability of the text. This gives us the opportunity to do not consider the readability constraint, one of the major obstacles in the text adversarial machine learning field.

Our goal was twofold: i) understand how different pipelines respond to ZeW attack, and ii) whether commercial applications are vulnerale to ZeW attack.
In Section~\ref{sec.res_indoor} we showed that different implementation are vulnerable with different magnitude to the attack, while character based models show promesing ``\textit{security by design}'' patterns. We then demonstrate the ferocity of the attack on commercial solutions (Section~\ref{sec.results}):
on 12 services developed by top IT companies such as Amazon, Google, IBM, and Microsoft, 11 show vulnerabilities. Among these 11, only 3 present a good resistance to the attack, while the remaining 8 are heavily affected.  
The simplicity of the attack allows it to spread to a broad population of malicious users and activities since no prior knowledge of machine learning theory is required. 

Potentially, we can find several use-cases of our attack and not only hate-speech manipulation. For example, we can consider web data mining techniques that can be used for counter-terrorism~\cite{Thuraisingham} where, NLP technologies can help to identify malicious content. In this scenario, terrorists could use ZeW to obfuscate the contents of their web-pages, affecting the performance of the analyzer. Because of this, our simple but effective countermeasure based on an input-validation technique should be integrated into every real-life NLP tool.

The security of machine learning applications is strictly related to their input domain. Computer Vision has different challenges compared to Natural Language Processing, which has different challenges compared to the signal domain. Moreover, state-of-the-art mainly focus on the security of the machine learning models, by forgetting that a machine learning application is composed by several stages where the ML model is only one of these. 
In conclusion, we believe that novel malicious opportunities can be derived by exploiting vulnerabilities of different components of the ML pipeline, and one of these directions is the leverage of multiple representations of the text, such as the usage of ASCII and UNICODE.